\documentclass[12pt,manuscript]{aastex}

\def\hii   {\protect\ion{H}{2}}
\def\nii   {\protect\ion{N}{2}}

\slugcomment{Accepted for publication by The Astronomical Journal}

\shorttitle{NGC 1311 Hot Stars}
\shortauthors{Eskridge et al.}

\begin{document}

\title{The Young Stellar Population of the Nearby Late-Type Galaxy NGC 
1311\footnotemark[1]}

\author{Paul B.~Eskridge\altaffilmark{2,3}}
\email{paul.eskridge@mnsu.edu}

\author{Rogier A.~Windhorst\altaffilmark{3}, Violet 
A.~Mager\altaffilmark{4,5}, \& Rolf A.~Jansen\altaffilmark{3}}

\footnotetext[1]{Based on observations with the NASA/ESA Hubble Space 
Telescope, obtained at the Space Telescope Science Institute, which is operated 
by the Association of Universities for Research in Astronomy, Inc.~under NASA 
contract No.~NAS5-26555.}
\altaffiltext{2}{Department of Physics \& Astronomy, Minnesota State
University, Mankato, MN 56001}
\altaffiltext{3}{School of Earth \& Space Exploration, Arizona State 
University, Tempe, AZ 85287}
\altaffiltext{4}{Department of Physics \& Astronomy, Arizona State University,
Tempe, AZ 85287}
\altaffiltext{5}{Observatories of the Carnegie Institution of Washington,
Pasadena, CA 91101}

\begin{abstract}
We have extracted PSF-fitted stellar photometry from near-ultraviolet, optical 
and near-infrared images, obtained with the {\it Hubble Space Telescope}, of 
the nearby ($D \approx 5.5$ Mpc) SBm galaxy NGC 1311.  The ultraviolet and 
optical data reveal a population of hot main sequence stars with ages of 2--10
Myr.  We also find populations of blue supergiants with ages between 10 and 40 
Myr and red supergiants with ages between 10 and 100 Myr.  Our near-infrared 
data shows evidence of star formation going back $\sim$1 Gyr, in agreement with 
previous work.  Fits to isochrones indicate a metallicity of $Z \approx 0.004$. 
The ratio of blue to red supergiants is consistent with this metallicity.  This 
indicates that NGC 1311 follows the well-known luminosity-metallicity relation 
for late-type dwarf galaxies.  About half of the hot main sequence stars and 
blue supergiants are found in two regions in the inner part of NGC 1311.  These 
two regions are each about 200 pc across, and thus have crossing times roughly 
equal to the 10 Myr age we find for the dominant young population.  The 
Luminosity Functions of the supergiants indicate a slowly rising star
formation rate (of $\sim$10$^{-3}\,M_{\odot}~{\rm yr}^{-1}$) from $\sim$100 Myr 
ago until $\sim$15 Myr ago, followed by a strong enhancement (to 
$\sim$10$^{-2}\,M_{\odot}~{\rm yr}^{-1}$) at $\sim$10 Myr ago.  We see no 
compelling evidence for gaps in the star-forming history of NGC 1311 over the 
last 100 Myr, and, with lower significance, none over the last Gyr.  This 
argues against a bursting mode, and in favor of a gasping or breathing mode for 
the recent star-formation history.
\end{abstract}


\section{Introduction}

The original formulation of the stellar population concept 
(\citeauthor{baa}\citeyear{baa}) was driven by the realization that the ages 
and metallicities of Galactic stars were related to their kinematics.  While 
Baade's identification of the young, metal-rich disk stars as Pop I, and the 
old, metal-poor halo stars as Pop II was a crucial starting point, observations 
in the last half century have identified both old and metal-rich stellar 
populations associated with the inner parts of spheroids (e.g., 
\citeauthor{tfw}\citeyear{tfw}; \citeauthor{isf}\citeyear{isf}) and young and
metal-poor stellar populations associated with low-luminosity star-forming
galaxies (e.g., \citeauthor{m98}\citeyear{m98}; 
\citeauthor{wet}\citeyear{wet}) and with the outer parts of luminous disks 
(e.g., \citeauthor{zkh}\citeyear{zkh}; \citeauthor{mcb}\citeyear{mcb}; 
\citeauthor{kea}\citeyear{kea}).  Understanding young, metal-poor stellar 
populations is an essential task in modern astrophysics, both because it allows 
us a fuller understanding of star formation and stellar evolution as a function 
of metallicity, and because it offers us potential insights into the nature of 
star formation in the early, low-metallicity Universe.

Our understanding of stellar populations has received an enormous boost from 
our ability to obtain arcsecond (or better) resolution images in the 
ultraviolet (UV) and near infrared (NIR).  In the NIR, the last decade has 
shown the complimentary power of large ground-based surveys 
(\citeauthor{p03}\citeyear{p03}; \citeauthor{jhk}\citeyear{jhk}) and 
higher-resolution {\it Hubble Space Telescope} (HST) Near Infrared Camera and
MultiObject Spectrograph (NICMOS) data.  In the UV, we are now seeing the 
benefits of combining insights from the large, but moderate-resolution, GALEX 
(\citeauthor{glx}\citeyear{glx}) database with high-resolution imaging from HST 
(e.g., \citeauthor{ozp}\citeyear{ozp}).  Imaging at HST resolution provides the
opportunity to study resolved stellar populations out to distances of $\sim$10 
Mpc.  Broad wavelength coverage (UV through NIR) allows us to investigate the 
full history of star formation in nearby galaxies.

NGC 1311 is a very nearby ($D \approx 5.5$ Mpc), but little-studied late-type 
(SBm) galaxy.  \citeauthor{tully}(\citeyear{tully}) identify it as a member of 
the 14$+$14 Association, a loose group dominated by the luminous spiral NGC 
1313.  Table 1 summarizes the basic properties of the system.  NGC 1311 was a 
target in two broad-band HST snapshot surveys (GO programs 9124 and 9824; 
\citeauthor{wind02}\citeyear{wind02}, \citeauthor{vio}\citeyear{vio}, 
\citeauthor{vtm}\citeyear{vtm}, and \S 2 below).  As a result, a set of 
broad-band images spanning a wide wavelength interval (0.3---1.6$\mu$m) at 
sub-arcsecond resolution now exists for this galaxy.  As NGC 1311 is quite 
nearby, its bright star clusters and luminous individual stars are detected as 
discrete sources.  We can thus probe the spatially resolved star-formation 
history of NGC 1311 by studying the broad-band spectral energy distributions of 
its star clusters, its individual stars, and its unresolved light.  Our results 
for the star clusters are presented in \citeauthor{paper1}(\citeyear{paper1}).  
This paper is concerned with the individual luminous stars.  We shall address 
the unresolved light in a future publication.

In \S 2 we briefly summarize the HST observational data.  We present the 
observed properties of the luminous stars in \S 3, and our analysis of these
observations in \S 4.  We summarize our conclusions, and discuss issues for
further research in \S 5.

\section{Observational Data}

The data for this study are a set of UV, optical and NIR images obtained with
the Wide-Field and Planetary Camera 2 (WFPC2) and NICMOS on board HST.  We have 
WFPC2 images taken through the F300W, F606W and F814W filters, and a NICMOS 
image taken with the NIC3 camera through the F160W filter.  A summary of the 
observations is given in Table 2.  The F300W image was obtained as part of the 
HST program GO-9124 ``Mid-UV Snapshot Survey of Nearby Irregulars:  Galaxy 
Structure and Evolution Benchmark'' (R.~Windhorst PI).  For this program, the 
centers of the target galaxies were placed on the WF3 chip.  Details of the 
observing and reduction procedures for these data are given in
\citeauthor{wind02}(\citeyear{wind02}).  The F160W image was obtained as part
of the HST program GO-9824 ``NIC3 SNAPs of Nearby Galaxies Imaged in the
mid-UV:  The Remarkable Cool Stellar Population in Late-Type Galaxies''
(R.~Windhorst PI).  Details of the observing and reduction procedures for these
data are given in \citeauthor{vio}(\citeyear{vio}).  The archival WFPC2 F606W
and F814W images of NGC 1311 were obtained as part of the HST program GO-9162 
``Local Galaxy Flows and the Local Mass Density'' (R.~Tully PI).  The WFPC2 
Wide-Field Camera (WFC) spatial sampling is $\approx 0{''}\llap.10$ per pixel.  
The Planetary Camera (PC) spatial sampling is $\approx 0{''}\llap.05$ per 
pixel.  No stars are detected in the F300W PC image, but we do use data from
the PC in F606W and F814W.  The NIC3 spatial sampling is $0{''}\llap.20$ per 
pixel.  We show images of the central 42$'' \times 26''$ of NGC 1311 in the 
four observed bands in Figure 1.

\section{Stellar Photometry}

We used HSTPhot\footnotemark[6] (\citeauthor{hstpht}\citeyear{hstpht}) to
extract PSF-fitted stellar photometry from the WFPC2 images.  HSTPhot is 
designed to perform stellar photometry on WFPC2 images, including aperture 
corrections, charge-transfer efficiency (CTE) corrections, and zero-point 
calibrations.  The zero-points to the VEGAMAG system are thus updated from 
those of \citeauthor{holtz}(\citeyear{holtz}).  For the NICMOS image, we first 
applied the non-linearity correction determined by
\citeauthor{nonlin}(\citeyear{nonlin}).  This refines the zero-point
calibration from the 2004 June standard (\citeauthor{nic70}\citeyear{nic70}).  
We then extracted stellar photometry using the version of DAOPHOT
(\citeauthor{hat87}\citeyear{hat87}) embedded in the XVISTA image analysis
package (\citeauthor{stov88}\citeyear{stov88}).  For the NICMOS photometry, we
had to determine the aperture correction manually by measuring the asymptotic
count rates for two bright, isolated stars in the observed NIC3 image
mosaic.  The correction from a 2-pixel radius to infinite aperture is 
0.10$\pm$0.02 magnitudes.  All of our photometry is calibrated to the VEGAMAG 
system.  We detect a total of 4369 stars in at least two bands.  Table 3 shows 
our photometry for the first ten objects in the sample.  The full table is 
presented in the on-line edition of this paper.  Column 1 gives a running
ID number (in order of increasing right ascension).  Columns 2 and 3 give the 
J2000.0 Right Ascension and Declination determined from the WFPC2 WCS header
information.  Columns 4 and 7 give the WFC chip on which the star is found in 
the F300W and F606W/F814W observations, respectively.  Columns 5 and 6 give the 
$X$ and $Y$ pixel positions of the star in the F300W image, column 8 and 9 give 
the F606W/F814W pixel positions, and column 10 and 11 give the NICMOS pixel 
positions.  Columns 12 through 15 give the measured magnitudes (top row) and 
errors (bottom row) in F300W ($UV_{300}$), F606W ($V_{606}$), F814W ($I_{814}$) 
and F160W ($H_{160}$).  Columns 16 through 21 give the measured colors (top 
row) and errors (bottom row) in $(UV_{300} - V_{606})$, $(UV_{300} - I_{814})$, 
$(UV_{300} - H_{160})$, $(V_{606} - I_{814})$, $(V_{606} - H_{160})$, and 
$(I_{814} - H_{160})$ respectively.

\footnotetext[6]{We used the May 2003 revision of HSTPhot v.~1.1.5b obtained
from {\tt http://purcell.as.arizona.edu/wfpc2\_calib/}.}

\subsection{Completeness Tests}

We conducted a series of artificial star tests on our images to determine both
the completeness and accuracy of our photometry.  For the WFPC2 images, we
used the artificial star capability of HSTPhot.  For the NIC3 image, we used
the version of DAOPHOT embedded in the XVISTA image analysis package.  Figure
2 and Table 4 show our completeness fraction as a function of magnitude in all 
four bands.  Our photometric accuracy is shown in Figure 3.  Table 5 shows the 
mean offset and dispersion as a function of magnitude in all four bands.  Our
50\% completeness limits, in magnitudes, are 22.2 (F300W), 25.7 (F606W), 24.4
(F814W) and 20.2 (F160W).  At these limits, the artificial star tests have
offsets that are statistically equivalent to zero, and dispersions of 0.1 to 
0.2 mag.

\subsection{Color-Magnitude and Color-Color Diagrams}

Figure 4 shows a suite of color-magnitude diagrams (CMDs) for the resolved 
stellar populations of NGC 1311.  The long-dashed lines show the 50\% 
completeness levels for the relevant bands.  The star clusters studied by 
\citeauthor{paper1}(\citeyear{paper1}) are not plotted.  The $V_{606}$ 
vs.~$(V_{606} - I_{814})$ CMD shows two plumes of luminous stars, separated by 
about a magnitude in color.  The dashed boxes in Fig.~4a show the two regions 
of the CMD that we use to select the stars in these two plumes (the blue-plume 
and red-plume stars, hereafter).  

Inspection of Fig.~4 allows us to make several qualitative statements about the
resolved stellar populations.  First, the blue-plume stars appear to be a mix
of hot main-sequence (MS) stars and post-main sequence blue supergiants (BSGs), 
based on where they lie in Fig.~4b,c.  Second, only the turn-off stars of the 
hot main-sequence are bright enough in $I_{814}$ to be unambiguously detected.  
The main-sequence stars are bluer on average in $(UV_{300} - V_{606})$ than the 
BSGs (see Fig.~4b).  We use this to refine our classification of blue-plume
stars as either BSGs (more luminous and redder than the diagonal line in 
Fig.~4b) or MS stars (less luminous and bluer than this line).  Third, the 
red-plume stars are red supergiants (RSGs).  They are nearly all undetected in 
$UV_{300}$, but show a clear vertical plume structure in Fig.~4d.  In 
Fig.~4bcd, the BSGs are circled and the RSGs are crossed.  We note that the tip 
of the red giant branch (RGB) of an old stellar population is at an absolute 
magnitude of $M_I \approx -4.0$, which corresponds to an apparent magnitude of 
$m_I \approx 24.7$ for NGC 1311.  This is much fainter than the magnitudes of 
the red-plume stars in Fig.~4.

We show a selection of color-color diagrams in Figure 5, chosen to show both
the BSGs and RSGs.  In all cases, the BSGs are well isolated from the other 
detections, and the RSGs are mixed with red stars from neither plume.  These 
are the faint red stars in Fig.~4d, and appear to be evolved stars from 
populations older than the bright RSG stars.   As we have data in a somewhat 
unusual set of filters (even for HST observations!), it seems prudent to 
compare our data to stellar population models before attempting any more 
detailed interpretation.

\subsection{Isochrone Matching}

We compare our stellar photometry to isochrones derived from the models of
\citeauthor{thiso}(\citeyear{thiso}).  In order to plot the isochrones with the
photometry, we adopt the distance estimate for NGC 1311 of 
\citeauthor{tully}(\citeyear{tully}), who quote $D = 5.45 \pm 0.08$ Mpc based 
on the magnitude of the tip of the red giant branch.  There appear to be no
metallicity measurements of NGC 1311 in the literature, although 
\citeauthor{haloc}(\citeyear{haloc}) report [\nii] / H$\alpha = 0.08$, 
consistent with a moderately low metallicity.  For the adopted distance, we 
find that the isochrones with $Z=0.004$ provide the best fit to the stellar 
photometry.  The more metal-poor isochrones have main sequences that are 
substantially bluer than observed in all available colors.  The isochrones more 
metal-rich than $Z=0.008$ do not predict stars luminous enough in the UV or
blue enough in colors including F300W compared to the data.  While crowding 
and/or binarity could artificially inflate the brightness of some fraction of 
our stars, our artificial star tests indicate that this is not a significant 
problem at the bright end of our photometry.  We adopt a metallicity of 
$Z=0.004$ below.

In Figures 6\&7, we show the suite of CMDs and color-color diagrams from 
Figs.~4\&5, with the \citeauthor{thiso}(\citeyear{thiso}) $Z=0.004$ isochrones 
overlayed.  The most luminous stars in NGC 1311 have absolute visual magnitudes 
of about $-8.5$.  Based on the isochrone comparison, the CMDs show a mix of 
young ($\la$10 Myr old) MS stars, and supergiants with ages ranging from 
$\sim$10 Myr up to $\sim$1 Gyr\footnotemark[7].  There is some evidence of a 
small population of stars younger than 10 Myr (see Fig.~6b).  The isochrones 
are broadly consistent with the stellar distribution in the color-color 
diagrams, with a few exceptions.  There are stars, including a few of the 
BSGs, that appear anomolously red in colors involving $H_{160}$.  
Inspection of the images shows that the stars in question typically have faint 
companions visible in the F606W and F814W images.  At the poorer sampling of 
the F160W image, these companions are blended with the brighter stars.  If the 
companion stars are red, this will create exactly the offset we observe in the 
color-color plots.  There are a few stars, not selected as supergiants, that 
have colors that are very red in colors involving F606W and redder filters.  
These stars are too faint in $V_{606}$, given their magnitudes in $I_{814}$ and 
$H_{160}$, to match the isochrone models.  We speculate that these stars may be 
dust enshrouded asymptotic giant branch (AGB) stars.  There is also the 
possibility that a few objects are unresolved background Extremely Red Objects 
(e.g., \citeauthor{whf}\citeyear{whf}).

\footnotetext[7]{Our F606W and F814W images are deep enough to reveal only the 
tip of the old RGB that is clearly revealed in the ACS data discussed by 
\cite{tully}.  As we are concerned with the young stellar population, this is
not a central issue for our study.}

\section{Discussion}

The isochrone matching reinforces the qualitative impressions from the CMDs
and color-color plots.  The blue-plume stars are a mix of hot MS stars with
ages of $\la$10 Myr and BSGs with ages from $\sim$10 up to $\sim$40 Myr.  The 
red-plume stars are red supergiants with ages from $\sim$10 up to $\sim$100 
Myr.  The hot main sequence stars are generally well-matched by the same 10 Myr 
isochrone that describes the youngest of the BSGs.  The distribution of MS and
BSG stars in Fig.~6abc, as well as the $H\alpha$ emission reported by 
\cite{haloc}, are evidence for star formation in the last $\sim$10 Myr, but at 
a rate much lower than 10 Myr ago.  We thus see evidence for a burst of star 
formation about 10 Myr ago superposed on what had been lower-level, but ongoing 
star formation over (at least) the previous 100 Myr or so.  

\subsection{Blue and Red Plume Stars}

The use of the ratio of blue to red supergiants (B/R ratio hereafter) as an 
astrophysical diagnostic of stellar populations was first suggested by
\citeauthor{vdb}(\citeyear{vdb}).  The B/R ratio turns out to depend on
(at least) the metallicity and age of a stellar population.  
\citeauthor{mnm}(\citeyear{mnm}) and \citeauthor{emm}(\citeyear{emm}) provide 
recent work on the relevant theory, while \citeauthor{dps}(\citeyear{dps}) and 
\citeauthor{umm}(\citeyear{umm}) present recent observational results in the 
context of nearby low-luminosity star-forming galaxies.

The boxes in Fig.~4a enclose 280 blue-plume and 190 red-plume stars (295 and 
200, after correcting for incompleteness).  However, as noted in \S 3.2, not 
all the blue-plume stars identified in Fig.~4a are BSGs.  Also, as pointed out
by \citeauthor{dps}(\citeyear{dps}), the blue and red supergiant plumes in the
CMDs of nearby galaxies will, in general, be due to stars of a range of ages.
As, even for a fixed metallicity, the B/R ratio is a function of age, some care
is required to choose fair samples of stars to evaluate the B/R ratio.  We have 
small samples to work with, so we choose to define three age-based samples, 
using the isochrones in Fig.~6a, with rough ages of 10 Myr, 18 Myr and 32 Myr.  
It is well known that models of RSGs do not produce stars as red as are 
observed (see \citeauthor{umm}\citeyear{umm} and references therein).  Thus we 
include, as RSGs of a given age, those stars that lie to the red of the 
isochrone red loops at a given luminosity.  In Table 6 we show the resulting 
age-based samples of BSGs and RSGs, and the B/R ratios for these samples.  The 
quoted uncertainties in Table 6 are from Poisson statistics.  At a given age, 
we have somewhat smaller B/R ratios than \citeauthor{dps}(\citeyear{dps}) found 
for Sextans A.  This is consistent with a higher metallicity for NGC 1311 than 
for Sextans A.  The Sextans A metallicity, from both \hii\ region spectroscopy 
(\citeauthor{skh}\citeyear{skh}) and CMD fitting 
(\citeauthor{dps}\citeyear{dps}), is $Z \approx 0.001$, compared to our adopted
metallicity for NGC 1311 of $Z = 0.004$.  This conclusion is reinforced by
the [\nii] / H$\alpha$ ratios of the two galaxies reported by \cite{haloc}.

\subsection{Distribution of Recent Star Formation}

In addition to the BSG and RSG samples selected above, we define a sample of
hot main sequence (MS) stars by taking all stars fainter than $UV_{300} = 20.7$
mag, and blueward of a line extending from ($-$2.6, 20.7) to (0, 22.8) in 
Fig.~4b.  We show the distribution of all three samples (MS, BSG, RSG) on the 
F300W WF3$+$WF4 image in Figure 8.  We note that there are a handful of 
sources that are not assigned to any of the three groups.  The brightest of 
these are the star clusters discussed in \cite{paper1}.  The fainter sources 
are only detected in $UV_{300}$.  These are likely to be MS stars, but as we do 
not have colors for them, we do not classify them as such.  There are two 
regions with strong concentrations of MS and BSG stars.  The boxes in Fig.~1a 
enclose these two regions.  In Figure 9, we show CMDs of the stars in the boxes 
and in the rest of the galaxy.  Nearly half the recent star formation in NGC 
1311 has occured in these two small regions, with the remainder broadly 
distributed throughout the system.  The total area of the two regions is about 
0.15 kpc$^2$ for our adopted distance, or about 2\% of the surface area of NGC 
1311.  These regions are also the locations of the strong H$\alpha$ emission
shown in \cite{hahi}.

The regions of enhanced recent star formation are about 10$''$ across.  This
corresponds to roughly 200 pc for our adopted distance.  Velocity-dispersion
measurements in late-type galaxies, for either stars or gas, are sparse but 
have characteristic values of about 10 to 20 km~s$^{-1}$ (e.g., 
\citeauthor{hew}\citeyear{hew}; \citeyear{dah}; \citeyear{dh2}).  If we assume 
that the velocity dispersion of the stars in the concentrations is similar to 
this, the stars will travel roughly 100 to 200 pc in 10 Myr.  Thus the sizes of 
the concentrations of young stars are consistent with the ages implied by the 
CMDs.  We can also compare these regions to Galactic OB associations.  OB 
associations are unbound groups of young stars that appear to be dispersing 
with internal velocity disperions of $\approx$5 km~s$^{-1}$ (e.g., 
\citeauthor{b64}\citeyear{b64}; \citeauthor{h80}\citeyear{h80}; 
\citeauthor{bb7}\citeyear{bb7}).  This gives a slightly longer lifetime, but
one that is still in range of 10s of Myr.

\subsection{Star Formation History}

The distribution of star cluster ages in NGC 1311 is consistent with a bursting 
mode of cluster formation, with active episodes of age $\sim$10 Myr, $\sim$100 
Myr and $\ga$1 Gyr (\citeauthor{paper1}\citeyear{paper1}).  The resolved stars
tell us a complimentary, but different story:  they do not show any compelling 
evidence for substantial periods of quiescence in the last $\sim$100 Myr.  Our 
CMDs show a clear population of stars with ages $\sim$10 Myr.  These are the MS 
and BSG stars seen in all panels of Fig.~6.  There is evidence of low-level 
on-going star formation over the last 10 Myr (see Fig.~6b). The distribution of 
BSGs and RSGs in the CMDs tell us that there was on-going star formation in NGC 
1311 from $\sim$10 Myr ago back to at least $\sim$100 Myr ago.  Although we are 
mainly concerned with the young stellar populations, we note that Fig.~6d and 
the work of \cite{tully} argue that star formation occured in a more or less 
steady fashion back to ages of at least $\sim$1 Gyr.  

We can make a somewhat more quantitative assessment of the recent star 
formation history from the luminosity function (LF) of the core Helium burning 
stars, as was first pointed out by \cite{sexa}.  In Figure 10, we show the BSG 
differential LFs of the BSG and RSG stars, with age points from the models of 
\cite{pad08}.  To convert the LFs into a star formation history we need to 
specify an initial mass function (IMF) and a stellar evolution model.  We use 
the Salpeter IMF \cite{imf} and the Padua stellar evolution models 
(\citeauthor{pad08}\citeyear{pad08}), and apply them following the precepts of
\cite{sexa}.  Figure 11 shows the resulting star formation history in the
age range 10--40 Myr from the BSGs.  The error bars in Fig.~11 represent 
statistical uncertainties from the star-counts only.  There are additional 
systematic uncertainties due to the adopted IMF and stellar evolution model.  
These will affect the absolute normalization of the star formation history, but 
should not substantially affect either its shape or its timescale.  The most 
luminous BSGs have ages of $\sim$10 Myr.  At ages above $\sim$40 Myr, it 
becomes impossible to isolate the BSGs on the CMDs.  Over the time period we 
can address, the rate of star formation appears to have been rising steadily, 
ending in a strong enhancement $\sim$10 Myr ago that corresponds to the most 
recent epoch of cluster formation.  

There are no BSGs more luminous than $M_{V606} = -8.5$.  This indicates that
star formation did not continue at the rate seen $\sim$10 Myr ago.  In 
principle, this is quantifiable by studying the MS LF.  The upper MS stars are 
most easily isolated in the $UV_{300}$ vs.~($UV_{300} - V_{606})$ CMD (see 
Figs.~4b, 6b).  The 50\% completeness limit is at $UV_{300} \approx 22.1$ mag,
corresponding to an absolute magnitude of $M_{UV300} \approx -6.6$.  This means
our data only probe the upper $\sim$1 magnitude of the MS LF.  We applied the 
iterative technique of \cite{sexa} to the completeness-corrected MS LF of NGC 
1311, again assuming a Salpeter IMF and the Padua models.  The result is that 
the upper MS stars provide evidence for star formation from 2 to 2.5 Myr ago.
The absolute star formation rate has large statistical uncertainty due to the
small number of stars and the large completeness corrections.  But there has
clearly been residual ongoing star formation in NGC 1311 since the enhancement
in star formation $\sim$10 Myr ago.  

We can extend the star formation history back to $\sim$100 Myr by taking 
advantage of the RSGs.  We caution the reader that theoretical modelling of the 
evolution of RSGs is on a less firm footing than that of BSGs 
(\citeauthor{lhlm}\citeyear{lhlm}).  Our resulting star formation history will 
therefore be subject to additional systematic uncertainty for ages larger than 
40 Myr.  However, we procede, again using a Salpeter IMF and the Padua models.  
The resulting star formation history is shown in logarithmic form in Figure 12. 
Over the last $\sim$100 Myr, the star formation rate in NGC 1311 appears to 
have been gradually increasing, peaking with the enhancement $\sim$10 Myr ago.  
We see no evidence for any gap in star formation over the last $\sim$100 Myr.  
These data are more consistent with a star formation history described as 
gasping or breathing (e.g., \citeauthor{mtg}\citeyear{mtg}; 
\citeauthor{evr}\citeyear{evr}; \citeauthor{sdq}\citeyear{sdq}) than bursting.

In \cite{paper1} we identified a small population of star clusters in NGC 1311.
Eight of these clusters have sufficient multiwavelength photometry to use for
age estimation.  These ages break into three groups.  Two clusters have ages
of $\la$10 Myr.  These clusters would appear to be associated with the
enhanced episode of star formation $\sim$10 Myr ago that is revealed by the 
BSGs (see Fig.~11).  We note that the total mass in stars formed in the 10 Myr
enhancement is about an order of magnitude larger than the mass estimated by
\cite{paper1} for the two young clusters,

There are no clusters with ages in the interval from 10 to 50 Myr.  The BSGs 
show that there was low-level on-going star formation throughout this interval, 
but no bound clusters survive from it.  As we note in \S 4.2, the BSGs are 
concentrated in two regions with properties similar to Galactic OB 
associations.  That is, the stars in that age range are spatially clumped, but 
they do not inhabit gravitationally bound clusters.  

Four clusters have ages between 50 Myr and 130 Myr.  The RSGs show evidence for 
a gradually increasing star formation rate toward the younger/more recent part 
of this range.  How this relates to the cluster formation in this epoch may be
discernable by a more extensive analysis of the HST Advanced Camera for
Surveys (ACS) data than was reported by \cite{tully}.  However, we note that
the total mass in intermediate-age clusters 
(\citeauthor{paper1}\citeyear{paper1}) is roughly equal to the total mass in
intermediate-age field stars based on the RSG LF.  Thus the star formation of
$\sim$100 Myr ago was efficient at producing clusters dense enough to remain 
bound over $\sim$100 Myr timescales, whereas the 10 Myr enhancement appears
to have produced mainly unbound stellar associations.

\subsection{The Luminosity-Metallicity Relation}

Fits to isochrones indicate a metallicity of $Z \approx 0.004$ ($[Z/H] \approx 
-0.7$), consistent with [\nii]/H$\alpha$ = 0.08 
(\citeauthor{haloc}\citeyear{haloc}).  Given its absolute magnitude (see Table 
1), this places NGC 1311 in the midst of the magnitude-metallicity relation for 
late-type dwarf galaxies (e.g., \citeauthor{m98}\citeyear{m98}).  We show this 
in Figure 13, where we plot B-band absolute magnitudes and nebular oxygen 
abudances for star forming dwarf galaxies from a number of recent studies 
(\citeauthor{scm}\citeyear{scm}; \citeauthor{l03a}\citeyear{l03a}; 
\citeyear{l03b}; \citeyear{l05}; \citeyear{l06}; \citeyear{l07}; 
\citeauthor{lns}\citeyear{lns}).  To place NGC 1311 on the figure, we convert 
the linear metallicity ($Z = 0.004$) to logarithmic notation, assuming 
$Z_{\odot} = 0.019$.  This gives $[Z/H] = -0.68$.  If we further assume that 
$[Z/O] = 0$ for NGC 1311, and take $12 + \log(O/H)_{\odot} = 8.93$ from 
\cite{ag89}, this gives us an estimate of $12 + \log(O/H) \approx 8.25$ for NGC 
1311.  This is shown in Fig.~10 as a large diamond.  As noted in \S 4.1, our 
metallicity estimate is also consistent with the observed B/R ratios for 
supergiants as sorted by age.  

\section{Summary and Conclusions}

NGC 1311 is a nearby ($D \approx 5.5$ Mpc) late-type (SBm) galaxy.  We have 
analysed stellar photometry from HST images of this galaxy covering a 
wavelength range from 0.3---1.6$\mu$m.  We detect hot MS stars in the UV and
optical images, providing strong evidence for star formation as recently as
$\sim$2 Myr ago.  We also detect BSG and RSG stars that provide evidence of
ongoing star formation over the last $\sim$100 Myr.  Our NIR data reveal 
populations of stars as old as $\ga$1 Gyr, in agreement with the work of
\cite{tully}.  The field stars of NGC 1311 show a less burst-like age
distribution than do the star clusters in NGC 1311 
(\citeauthor{paper1}\citeyear{paper1}).  About half of the recent star 
formation, as traced by the MS and BSG stars, is confined to two regions about 
200 pc across in the bright central part of the system.  Assuming a velocity 
dispersion for the stars that is typical of Galactic OB associations, the 
crossing time is on the order of the age of 10 Myr determined from the CMDs. 
The metallicity of $Z \approx 0.004$ that follows from the isochrone fits and
the B/R ratio places NGC 1311 in the midst of the metallicity-luminosity
relation for low-luminosity late-type galaxies.

Our results for NGC 1311 are consistent with those found for the general 
population of late-type, low-luminosity galaxies (e.g., 
\citeauthor{mtg}\citeyear{mtg}; \citeauthor{m98}\citeyear{m98}; 
\citeauthor{hew}\citeyear{hew}; \citeauthor{yvl}\citeyear{yvl}; 
\citeauthor{bag}\citeyear{bag}; \citeauthor{sdq}\citeyear{sdq}; 
\citeauthor{sih}\citeyear{sih}; \citeauthor{skd}\citeyear{skd}; 
\citeauthor{wet}\citeyear{wet}).  The arising consensus is that such galaxies 
form stars over their entire histories.  The rate of star formation may change
with time, but only rarely does it cease entirely.  The long-term star forming 
histories of low-luminosity galaxies are best studied with deep long-wavelength 
optical ($RI$) or NIR ($JHK$) images.  Our UV and optical data allow us to 
explore the star forming history of NGC 1311 over the last $\sim$100 Myr.

Our next step is a study of the unresolved light of NGC 1311 with our combined 
WFPC2/NICMOS HST data.  As we have observations in four wavelengths from 
3000\AA\ to 1.6$\micron$, we will sample stellar populations from ages of 
$\sim$10 Myr to $\ga$10 Gyr.  Nearby star-forming galaxies are known to have 
unresolved UV light (e.g., \citeauthor{rlt}\citeyear{rlt}) at HST resolution.  
The stellar photometry we present in this paper provides information on 
individual MS stars as late as early B-type stars.  Late B-type and A-type 
stars will contribute significant diffuse UV light that traces the distribution 
of stars of $\sim$100 Myr ages.  The ancient stellar populations, dominating 
the NIR light, appear ubiquitous even in very late-type galaxies (e.g.,
\citeauthor{baa}\citeyear{baa}; \citeauthor{vio}\citeyear{vio}).  The 
combination of our cluster and stellar photometry with an analysis of the 
unresolved light should give us a clearer picture of the star formation history 
in this system, and a fuller understanding of the process of star formation in 
low-luminosity late-type galaxies in general.

\acknowledgments

PBE would like to thank the Department of Physics \& Astronomy at Minnesota
State University for support from a Faculty Research Grant during this project,
and to the School of Earth \& Space Exploration at Arizona State University for
their hospitality during his sabbatical visit.  We thank Andy Dolphin for
answering the first author's HSTPhot questions with patience.  This research
has made use of NASA's Astrophysics Data System Bibliographic Services, and the
NASA/IPAC Extragalactic Database (NED) which is operated by the Jet Propulsion
Laboratory, California Institute of Technology, under contract with the
National Aeronautics and Space Administration.  This work was supported in part 
by NASA Hubble Space Telescope grants HST-GO-09124* and HST-GO-09824* awarded 
by the Space Telescope Science Institute, which is operated by AURA for NASA 
under contract NAS 5-26555.

Facilities: \facility{HST}

{
\begin{deluxetable}{lcc}
\tablewidth{0pt}
\tablecolumns{3}
\tablecaption{Basic Properties}
\tablehead{
\colhead{} & \colhead{} & \colhead{References}}
\startdata
$m_B$ & 13.22$\pm$0.21 & 1 \\
$(B - V)$ & 0.46$\pm$0.02 & 1 \\
$\log{D_{25}}$ & 1.48$\pm$0.03 & 1 \\
$\log{R_{25}}$ & 0.59$\pm$0.05 & 1 \\
$V_{\odot}$ & 568$\pm$5 km/sec & 2 \\
$A_V$ & 0.07 & 3 \\
$E{(B-V)}$ & 0.021 & 3 \\
D & 5.45$\pm$0.08 Mpc & 4 \\
$M_B$ & $-$15.5 & 1,4 \\
$S_{HI}$ & 15.4 Jy\,{km/sec} & 2 \\
$M_{HI}/L_B$ & 0.46 $M_{\odot} / L_{\odot}$ & 1,2,4 \\
$EW(H\alpha +$[\nii]) & 33$\pm$2 \AA & 5 \\
$\lbrack$\nii$\rbrack / H\alpha$ & 0.08 & 5 \\
\enddata
\tablerefs{
1) de Vaucouleurs et al.~(1991);
2) Koribalski et al.~(2004);
3) Schlegel et al.~(1998);
4) Tully et al.~(2006)
5) Kennicutt et al.~(2008)
}
\end{deluxetable}
}

{
\begin{deluxetable}{cclcc}
\tablewidth{0pt}
\tablecolumns{5}
\tablecaption{Log of Observations}
\tablehead{
\colhead{Data Set} & \colhead{PID} & \colhead{Camera/Filter} & \colhead{Date} &
\colhead{Exposure} \\ \colhead{} & \colhead{} & \colhead{} & 
\colhead{dd-mm-yyyy} & \colhead{sec}}
\startdata
u6dw7101m &  9124 &  WFPC2/F300W & 27-08-2001 &  300     \\
u6dw7102m &  9124 &  WFPC2/F300W & 27-08-2001 &  300     \\
u6g22403m &  9162 &  WFPC2/F606W & 22-09-2001 &  300     \\
u6g22404m &  9162 &  WFPC2/F606W & 22-09-2001 &  300     \\
u6g22401m &  9162 &  WFPC2/F814W & 21-09-2001 &  300     \\
u6g22402m &  9162 &  WFPC2/F814W & 21-09-2001 &  300     \\
n8ou36010 &  9824 &  NIC3/F160W  & 28-11-2003 &  512     \\
\hline
\enddata
\end{deluxetable}
}

{
\begin{deluxetable}{cccccccccccrrrrcccccc}
\tabletypesize{\scriptsize}
\def\pb{\phantom{-}}
\rotate
\setlength{\tabcolsep}{0.02in}
\tablewidth{0pt}
\tablecolumns{21}
\tablecaption{Stellar Photometry}
\tablehead{
\colhead{ID} & \colhead{RA (J2000.0)} & \colhead{Dec.~(J2000.0)} & \colhead{chip}
& \colhead{X$_{300}$} & \colhead{Y$_{300}$} & \colhead{chip} & 
\colhead{X$_{V/I}$} & \colhead{Y$_{V/I}$} & \colhead{X$_{NIC}$} & 
\colhead{Y$_{NIC}$} & \colhead{$UV_{300}$} & \colhead{$V_{606}$} & 
\colhead{$I_{814}$} & \colhead{$H_{160}$} & \colhead{$UV_{300}$--$V_{606}$} & 
\colhead{$UV_{300}$--$I_{814}$} & \colhead{$UV_{300}$--$H_{160}$} & 
\colhead{$V_{606}$--$I_{814}$} & \colhead{$V_{606}$--$H_{160}$} & 
\colhead{$I_{814}$--$H_{160}$} \\
\colhead{} & \colhead{hh:mm:ss.sss} & \colhead{dd:mm:ss.ss} & \colhead{UV} & 
\colhead{pix} & \colhead{pix} & \colhead{V/I} & \colhead{pix} & 
\colhead{pix} & \colhead{pix} & \colhead{pix} & \colhead{mag} & \colhead{mag} & 
\colhead{mag} & \colhead{mag} & \colhead{mag} & \colhead{mag} & \colhead{mag} & 
\colhead{mag} & \colhead{mag} & \colhead{mag}}
\startdata
  1 & 3:20:00.637 & -52:12:00.83 &  &  &  & wf4 & 548.55 & 664.68 &        &        &        & 22.58 & 21.41 &        &        &        &       &  1.17 &        &        \\
    &              &               &     &        &        &     &        &        &        &        &        &  0.02 &  0.02 &        &        &        &       &  0.03 &        &        \\
  2 & 3:20:01.006 & -52:12:09.91 &  &  &  & wf4 & 645.57 & 650.82 &        &        &        & 22.03 & 20.93 &        &        &        &       &  1.11 &        &        \\
    &              &               &     &        &        &     &        &        &        &        &        &  0.02 &  0.02 &        &        &        &       &  0.02 &        &        \\
  3 & 3:20:02.103 & -52:12:04.99 &  &  &  & wf4 & 618.56 & 540.61 &        &        &        & 23.47 & 22.35 &        &        &        &       &  1.12 &        &        \\
    &              &               &     &        &        &     &        &        &        &        &        &  0.05 &  0.05 &        &        &        &       &  0.07 &        &        \\
  4 & 3:20:02.197 & -52:12:05.62 &  &  &  & wf4 & 626.68 & 533.60 &        &        &        & 23.18 & 22.08 &        &        &        &       &  1.10 &        &        \\
    &              &               &     &        &        &     &        &        &        &        &        &  0.03 &  0.03 &        &        &        &       &  0.04 &        &        \\
  5 & 3:20:02.265 & -52:12:03.02 &  &  &  & wf4 & 602.26 & 521.75 &        &        &        & 21.84 & 20.55 &        &        &        &       &  1.28 &        &        \\
    &              &               &     &        &        &     &        &        &        &        &        &  0.01 &  0.01 &        &        &        &       &  0.02 &        &        \\
  6 & 3:20:02.273 & -52:12:00.11 &  &  &  & wf4 & 573.71 & 515.00 &        &        &        & 22.47 & 21.31 &        &        &        &       &  1.16 &        &        \\
    &              &               &     &        &        &     &        &        &        &        &        &  0.02 &  0.02 &        &        &        &       &  0.03 &        &        \\
  7 & 3:20:02.284 & -52:12:01.41 &  &  &  & wf4 & 586.65 & 516.49 &        &        &        & 23.77 & 23.84  &        &       &        &        & $-$0.08 &        &        \\
    &              &               &     &        &        &     &        &        &        &        &       &  0.04 &  0.11  &        &        &        &       & $\pb$0.12 &       &        \\
  8 & 3:20:02.389 & -52:11:54.74 &  &  &  & wf4 & 525.16 & 489.27 &        &        &        & 24.51 & 23.92 &        &        &        &       &  0.60 &        &        \\
    &              &               &     &        &        &     &        &        &        &        &        &  0.08 &  0.14 &        &        &        &       &  0.16 &        &        \\
  9 & 3:20:02.405 & -52:11:59.42 &  &  &  & wf4 & 569.45 & 501.47 &        &        &        & 23.47 & 22.47 &        &        &        &       &  1.00 &        &        \\
    &              &               &     &        &        &     &        &        &        &        &        &  0.04 &  0.04 &        &        &        &       &  0.05 &        &        \\
 10 & 3:20:02.452 & -52:11:45.89 &  &  &  & wf4 & 437.16 & 468.12 &        &        &        & 23.48 & 22.53 &        &        &        &       &  0.95 &        &        \\
    &              &               &     &        &        &     &        &        &        &        &        &  0.04 &  0.04 &        &        &        &       &  0.06 &        &        \\
\hline
\enddata
\end{deluxetable}
}

{
\begin{deluxetable}{ccccc}
\tabletypesize{\small}
\tablewidth{0pt}
\tablecolumns{5}
\tablecaption{Completeness Fraction}
\tablehead{
\colhead{Magnitude} & \colhead{$UV_{300}$} & \colhead{$V_{606}$} & 
\colhead{$I_{814}$} & \colhead{$H_{160}$}}
\startdata
17.0 & & & & 0.99 \\
17.5 & & & 0.97 & 0.90 \\
18.0 & & & 0.97 & 0.92 \\
18.5 & 0.97 & & 0.97 & 0.91 \\
19.0 & 0.97 & 0.96 & 0.97 & 0.82 \\
19.5 & 0.97 & 0.97 & 0.97 & 0.65 \\
20.0 & 0.97 & 0.97 & 0.95 & 0.54 \\
20.5 & 0.96 & 0.97 & 0.96 & 0.37 \\
21.0 & 0.96 & 0.96 & 0.96 & 0.24 \\
21.5 & 0.92 & 0.96 & 0.97 & 0.16 \\
22.0 & 0.58 & 0.96 & 0.97 & \\
22.5 & 0.17 & 0.96 & 0.96 & \\
23.0 & 0.01 & 0.96 & 0.95 & \\
23.5 & & 0.96 & 0.94 & \\
24.0 & & 0.95 & 0.83 & \\
24.5 & & 0.94 & 0.38 & \\
25.0 & & 0.88 & 0.06 & \\
25.5 & & 0.61 & 0.01 & \\
26.0 & & 0.18 & & \\
26.5 & & 0.03 & & \\
27.0 & & 0.01 & & \\
\hline
\enddata
\end{deluxetable}
}

{
\begin{deluxetable}{ccccccccc}
\tabletypesize{\small}
\def\pb{\phantom{-}}
\tablewidth{0pt}
\tablecolumns{9}
\tablecaption{Artificial Star Photometric Accuracy}
\tablehead{
\colhead{Magnitude} & \colhead{$UV_{300}$} & \colhead{$\sigma_{300}$} & 
\colhead{$V_{606}$} & \colhead{$\sigma_{606}$} & \colhead{$I_{814}$} & 
\colhead{$\sigma_{814}$} & \colhead{$H_{160}$} & \colhead{$\sigma_{160}$}}
\startdata
17.0 & & & & & $-$0.028 & 0.230 & 0.008 & 0.014 \\
17.5 & & & & & $-$0.021 & 0.203 & 0.014 & 0.030 \\
18.0 & & & & & $-$0.004 & 0.075 & 0.021 & 0.037 \\
18.5 & & & $-$0.005 & 0.094 & $-$0.002 & 0.012 & 0.031 & 0.053 \\
19.0 & $-$0.008 & 0.025 & $-$0.001 & 0.010 & $-$0.002 & 0.013 & 0.060 & 0.115 \\
19.5 & $-$0.011 & 0.036 & $-$0.001 & 0.011 & $-$0.003 & 0.016 & 0.088 & 0.157 \\
20.0 & $-$0.015 & 0.043 & $-$0.001 & 0.013 & $-$0.005 & 0.022 & 0.120 & 0.202 \\
20.5 & $-$0.021 & 0.061 & $-$0.002 & 0.016 & $-$0.007 & 0.026 & 0.195 & 0.295 \\
21.0 & $-$0.027 & 0.081 & $-$0.003 & 0.017 & $-$0.010 & 0.034 & 0.268 & 0.426 \\
21.5 & $-$0.024 & 0.109 & $-$0.004 & 0.022 & $-$0.013 & 0.050 & 0.313 & 0.570 \\
22.0 & $-$0.009 & 0.144 & $-$0.008 & 0.033 & $-$0.022 & 0.061 \\
22.5 & $\pb$0.112 & 0.238 & $-$0.007 & 0.043 & $-$0.028 & 0.091 \\
23.0 & $\pb$0.358 & 0.463 & $-$0.011 & 0.066 & $-$0.044 & 0.144 \\
23.5 & $\pb$0.899 & 0.996 & $-$0.013 & 0.097 & $-$0.043 & 0.177 \\
24.0 & & & $-$0.025 & 0.120 & $\pb$0.054 & 0.216 \\
24.5 & & & $-$0.023 & 0.169 & $\pb$0.391 & 0.508 \\
25.0 & & & $\pb$0.010 & 0.217 \\
25.5 & & & $\pb$0.196 & 0.352 \\
26.0 & & & $\pb$0.728 & 0.871 \\
26.5 & & & $\pb$1.699 & 1.842 \\
27.0 & & & $\pb$2.524 & 2.645 \\
\hline
\enddata
\end{deluxetable}
}

{   
\begin{deluxetable}{ccccccc}
\tablewidth{0pt}
\tablecolumns{7}
\tablecaption{B/R Ratio}
\tablehead{
\colhead{Age} & \colhead{BSG \#} & \colhead{RSG \#} & \colhead{B/R Ratio} &
\colhead{BSG \#} & \colhead{RSG \#} & \colhead{B/R Ratio} \\
\colhead{Myr} & \colhead{uncorrected} & \colhead{uncorrected} & 
\colhead{uncorrected} & \colhead{corrected} & \colhead{corrected} & 
\colhead{corrected}}
\startdata
10 & 13$\pm$3.6 & 3$\pm$1.7 & 4.3$\pm$2.7 & 13.5$\pm$3.6 & 3.1$\pm$1.7 & 
4.4$\pm$2.7 \\
18 & 16$\pm$4.0 & 24$\pm$4.9 & 0.67$\pm$0.22 & 16.7$\pm$4.0 & 25.0$\pm$4.9 & 
0.70$\pm$0.22 \\
32 & 53$\pm$7.3 & 53$\pm$7.3 & 1.00$\pm$0.19 & 55.2$\pm$7.3 & 55.2$\pm$7.3 & 
1.00$\pm$0.19 \\
\hline
\enddata
\end{deluxetable}
}

\newpage

\figcaption{42$'' \times$26$''$ images of NGC 1311 in a) F300W, b) F606W, c) 
F814W, and d) F160W  The arrow in the upper-left corner of the F300W image 
points North, the attached line-segment points East.  The scale bar in the 
lower-right corner of the F300W image is 10$''$, and applies to all four 
panels.  The boxes in panel a) highlight regions discussed in \S 4.2.}

\figcaption{Recovery fraction as a function of input magnitude for artificial
stars in a) $UV_{300}$, b) $V_{606}$, c) $I_{814}$, and d) $H_{160}$.}

\figcaption{Magnitude difference versus input magnitudes for artificial
stars in a) $UV_{300}$, b) $V_{606}$, c) $I_{814}$, and d) $H_{160}$.  The 
y-axis shows input magnitude minus output magnitude.  The open circles with 
error bars show mean offsets in half-magnitude bins, as given in Table 5.}

\figcaption{Color-Magnitude Diagrams for the detected stars in NGC 1311.  The 
arrows in each plot show the foreground dereddening vectors.  a) $V_{606}$ 
vs.~$(V_{606} - I_{814})$, b) $UV_{300}$ vs.~($UV_{300} - V_{606})$, c) 
$UV_{300}$ vs.~($UV_{300} - I_{814})$, d) $V_{606}$ vs.~$(V_{606} - H_{160})$.  
The dotted boxes in a) show the selection for the blue-plume and red-plume 
stars.  We plot error bars for 1--5\% of the sample to avoid crowding.  In 
b,c,d the circled points are blue supergiant stars and the crossed points are 
red supergiant stars.  The long-dashed lines show the 50\% completeness limits. 
The solid line in b shows the separations between the blue supergiants and the 
hot main-sequence stars.}

\figcaption{Color-Color Diagrams for the detected stars in NGC 1311.  The
arrows in each plot show the foreground dereddening vectors.  a) $(V_{606} - 
I_{814})$ vs.~$(I_{814} - H_{160})$, b) $(V_{606} - H_{160})$ vs.~$(I_{814} - 
H_{160})$, c) $(V_{606} - I_{814})$ vs.~$(V_{606} - H_{160})$.  Symbols as in 
Fig.~4.}

\figcaption{The CMDs of Fig.~4 with isochrones from Girardi et al.~(2002).  The 
isochrones have Z=0.004, and cover a range in age from 4 Myr (blue) to 6 Gyr 
(red).  The solid lines show the 10 Myr, 100 Myr and 1 Gyr isochrones.  The 
dotted lines are younger, intervening, and older isochrones, sampled every 0.25 
dex in age.  Symbols as in Fig.~4}

\figcaption{The Color-Color Diagrams of Fig.~5 with isochrones from Girardi et 
al.~(2002).  The isochrones are those shown in Fig.~6.  Symbols as in Fig.~4.
The large arrows are reddening vectors for $A_V = 1$ mag.}

\figcaption{The distribution of MS (blue symbols), BSG (green symbols) and RSG
(red symbols) stars in NGC 1311 overplotted on the F300W WF3$+$WF4 image.}

\figcaption{CMDs of stars in the boxes shown in Fig.~1a (panels a and c) and
the rest of NGC 1311 (panels b and d).  In a and c, the stars from the eastern 
box are plotted in blue, and those from the western box are in red.  The 
isochrones and completeness lines are as in Fig.~6.  In b) and d), we plot 
error bars for 1--10\% of the sample to avoid crowding.}

\figcaption{Differential LFs of the BSG stars (top) and the RSG stars (bottom). 
The dotted histograms shows the observed counts.  The solid histograms are 
corrected for incompleteness.  The ages shown are from the models of Marigo et 
al.~(2008).}

\figcaption{Star formation history from 10 Myr to 40 Myr ago from the BSG 
stars.  The error bars are statistical errors from the observed counts.}

\figcaption{Star formation history from 10 Myr to 100 Myr ago from the BSG and 
RSG stars.  The star-formation rate is shown in logarithmic form.  The solid
and dashed histograms shows the star-formation history from the BSGs and RSGs
respectively.  The error bars are statistical errors from the observed counts.}

\figcaption{Absolute $B$-band magnitudes vs.~$12 + \log(O/H)$ for late-type
dwarf galaxies from several recent studies: solid circles -- Lee et al.~(2007); 
solid squares -- Lee et al.~(2003a); solid triangles -- Lee et al.~(2003b); 
open triangle (WLM) -- Lee et al.~(2005); open square (NGC 6822) -- Lee et 
al.~(2005); open circle (NGC 1705) -- Lee et al.~(2004); crosses -- Skillman
et al.~(2003).  The large diamond shows our estimate for NGC 1311.}

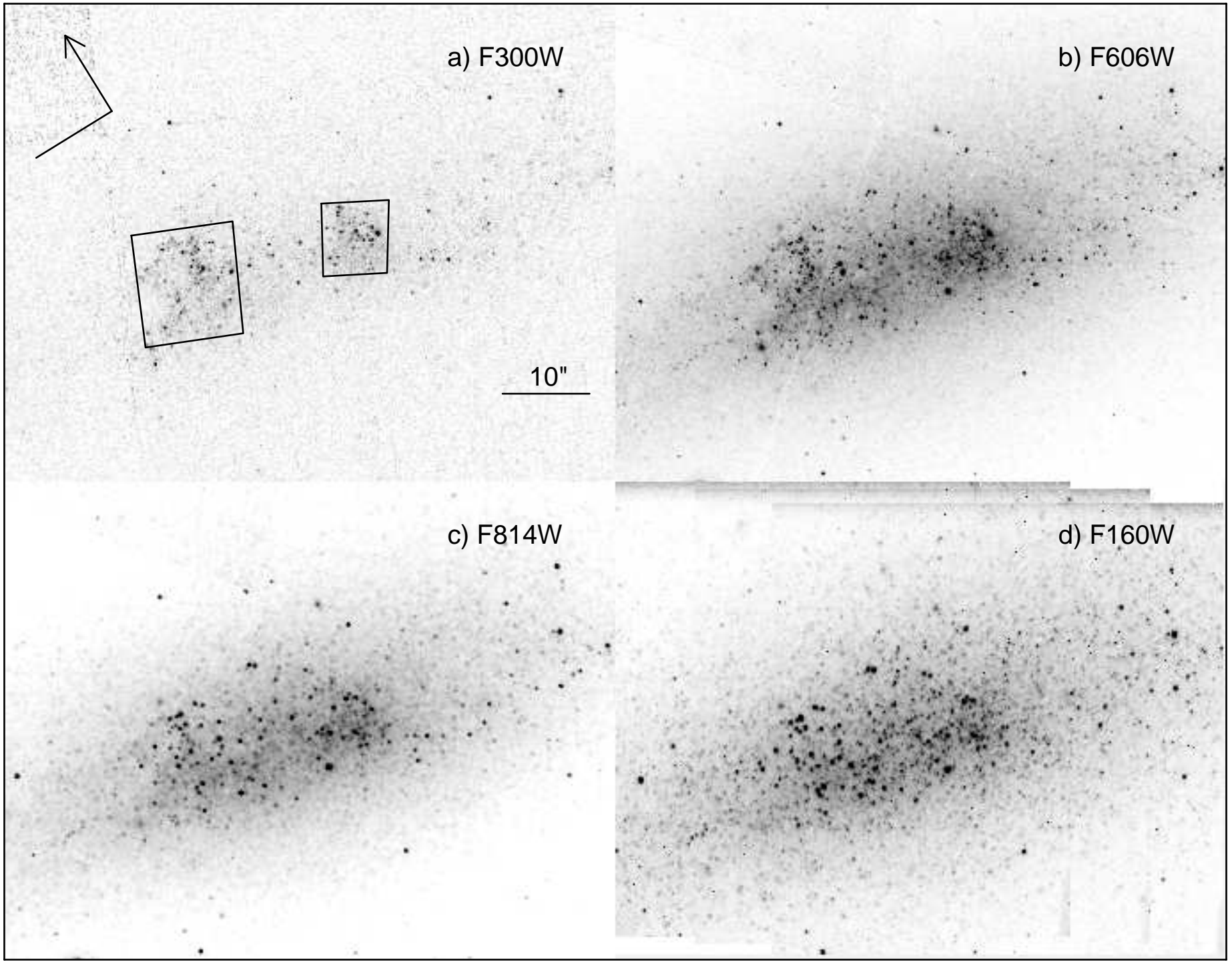
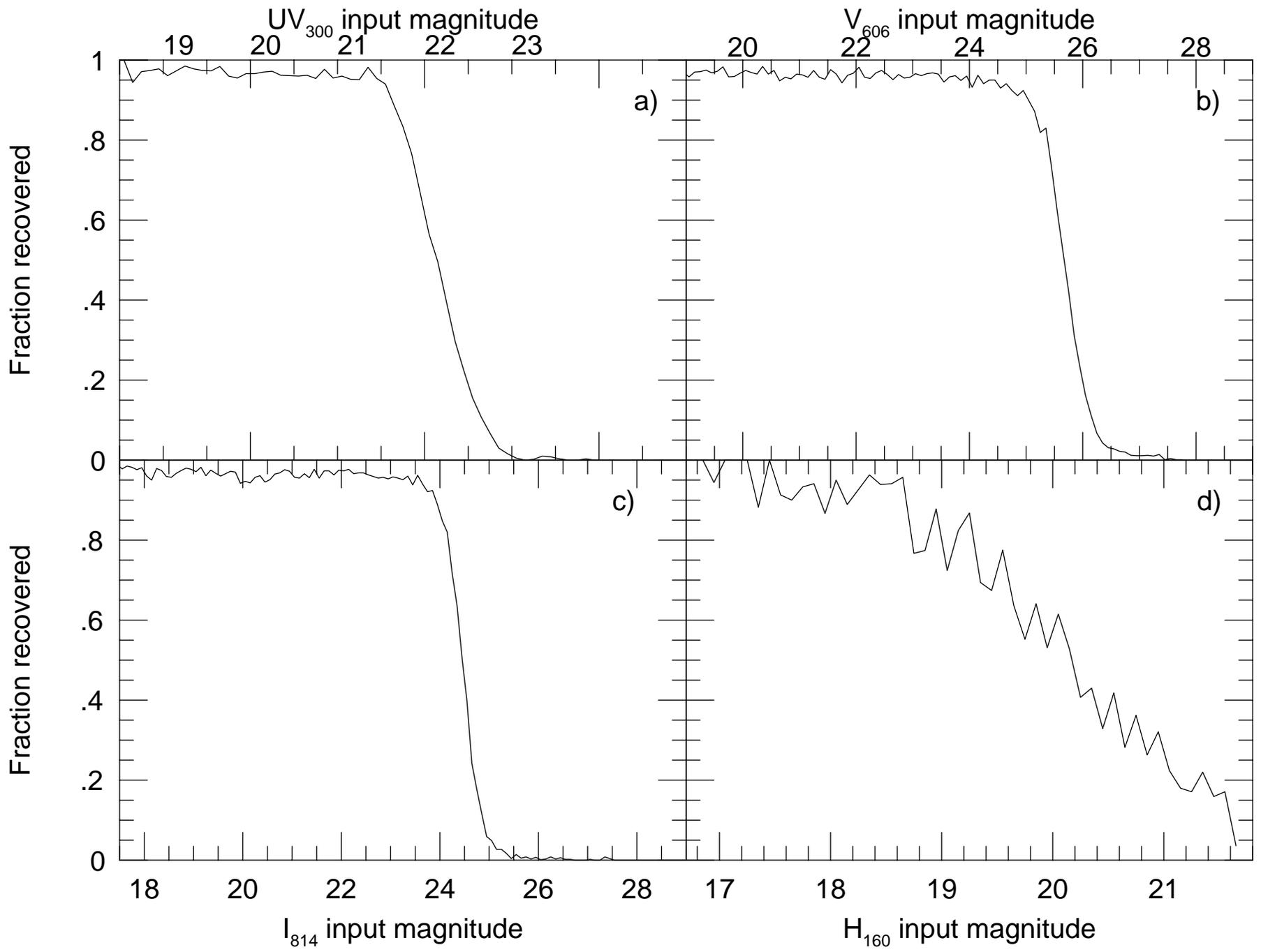
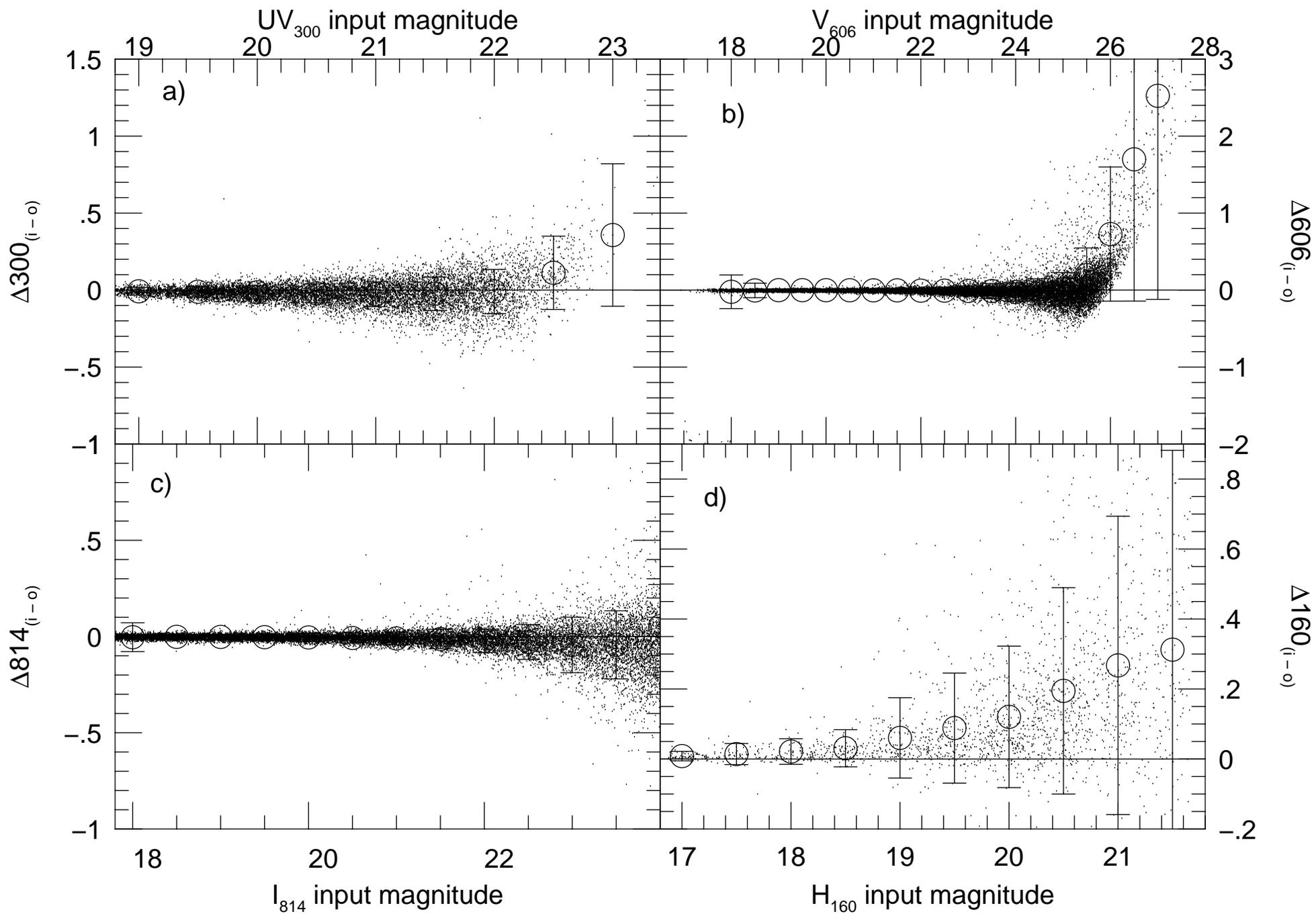
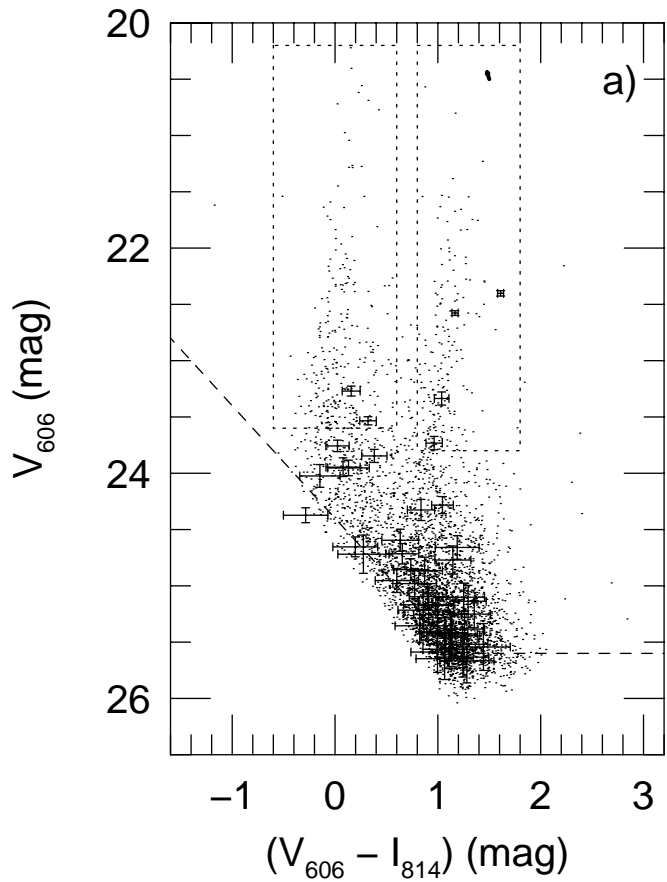
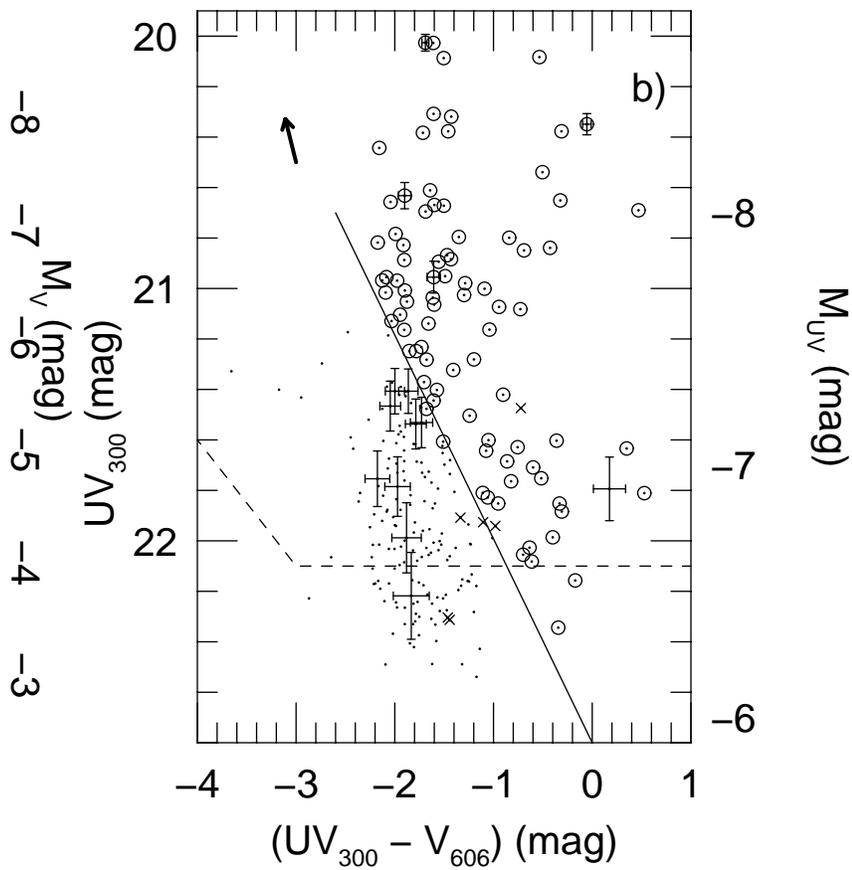
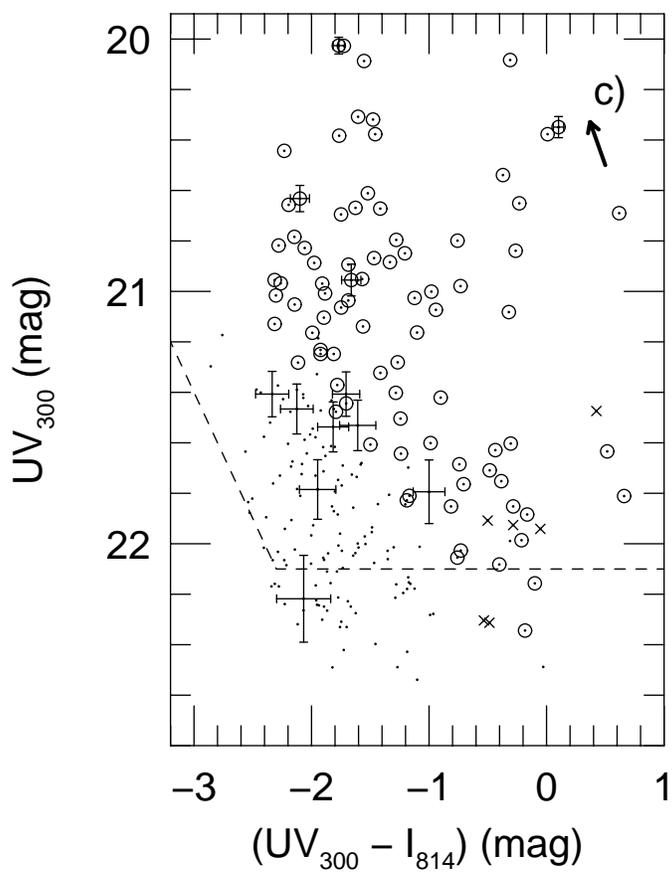
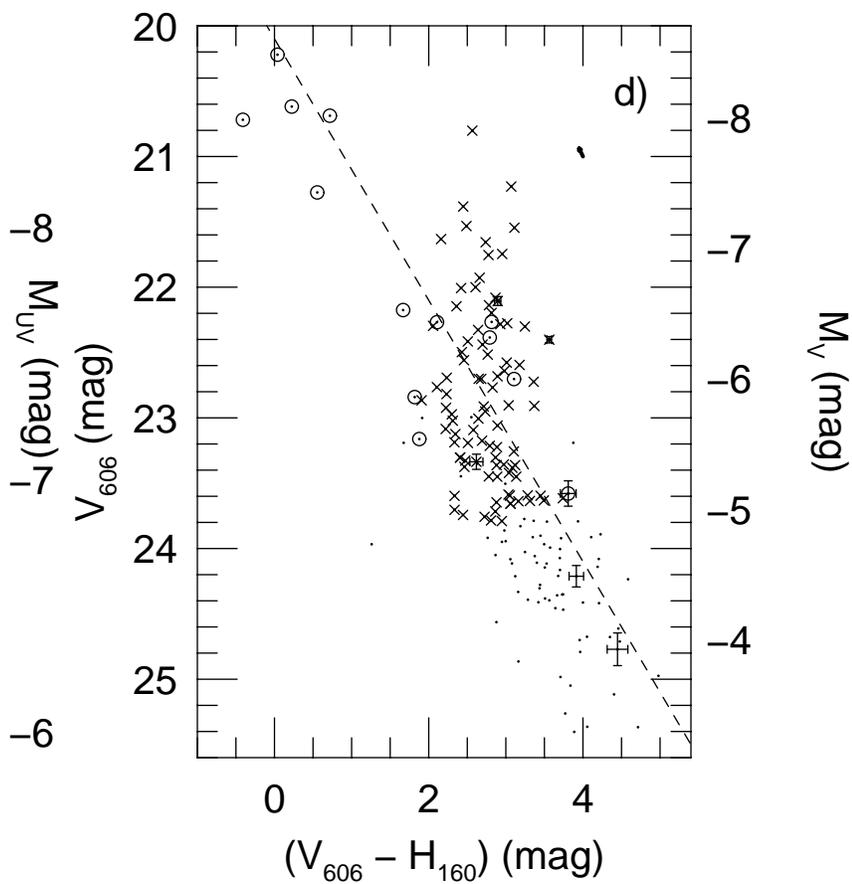
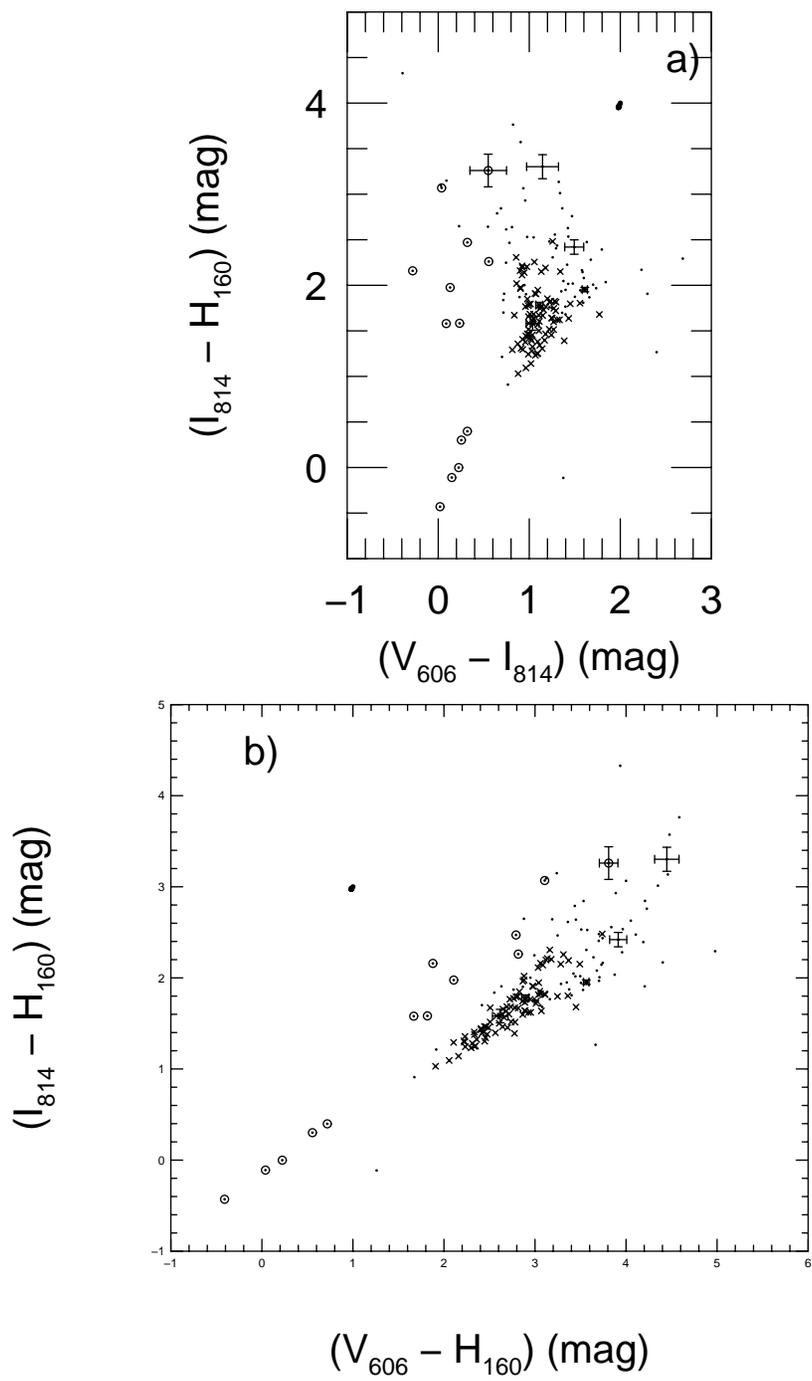
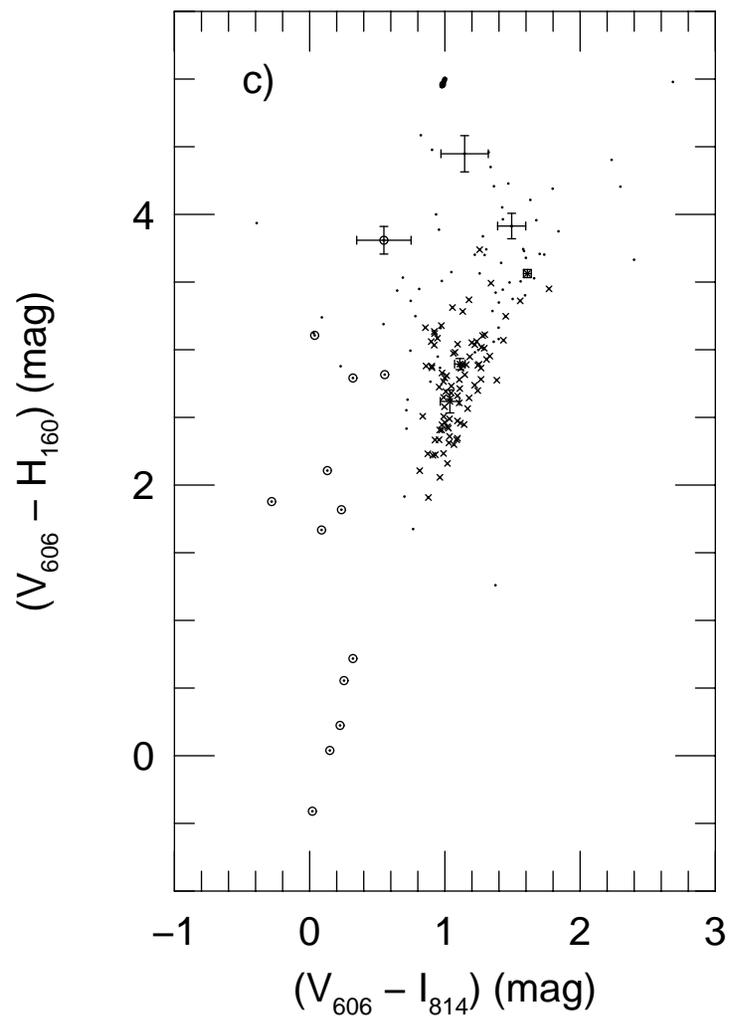
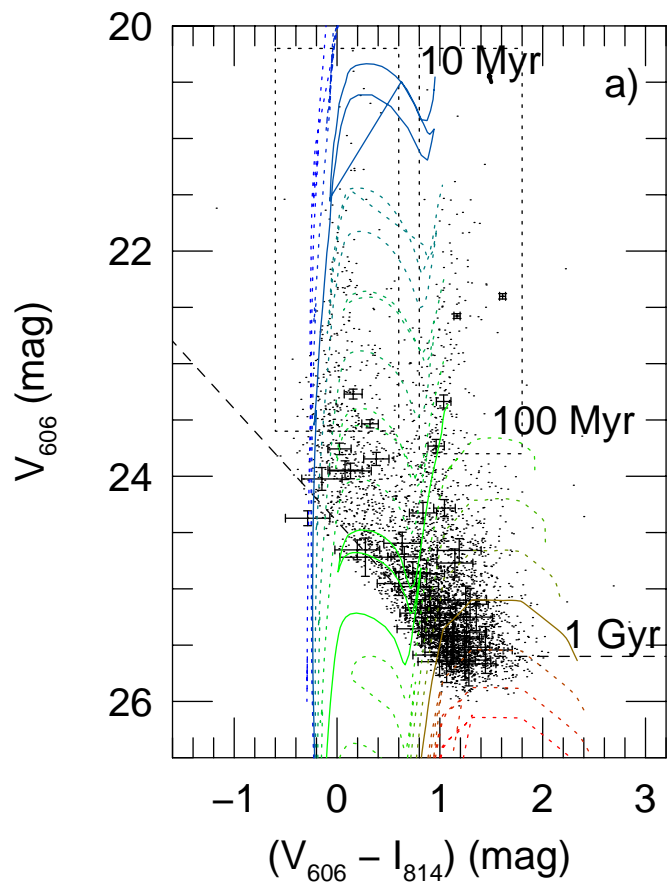
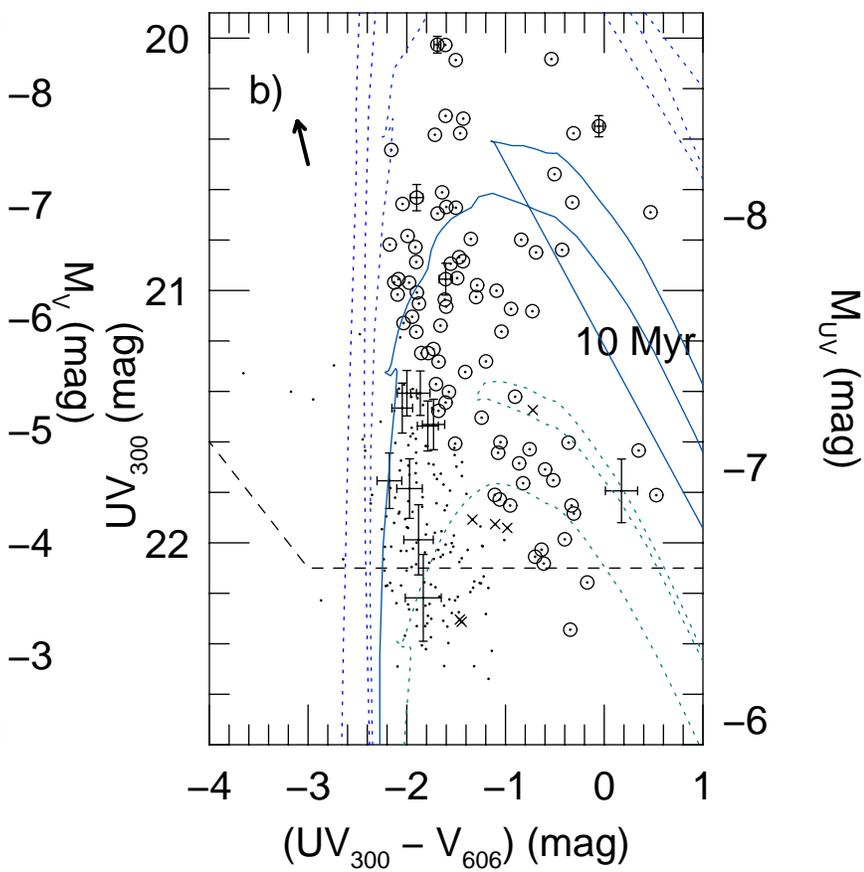
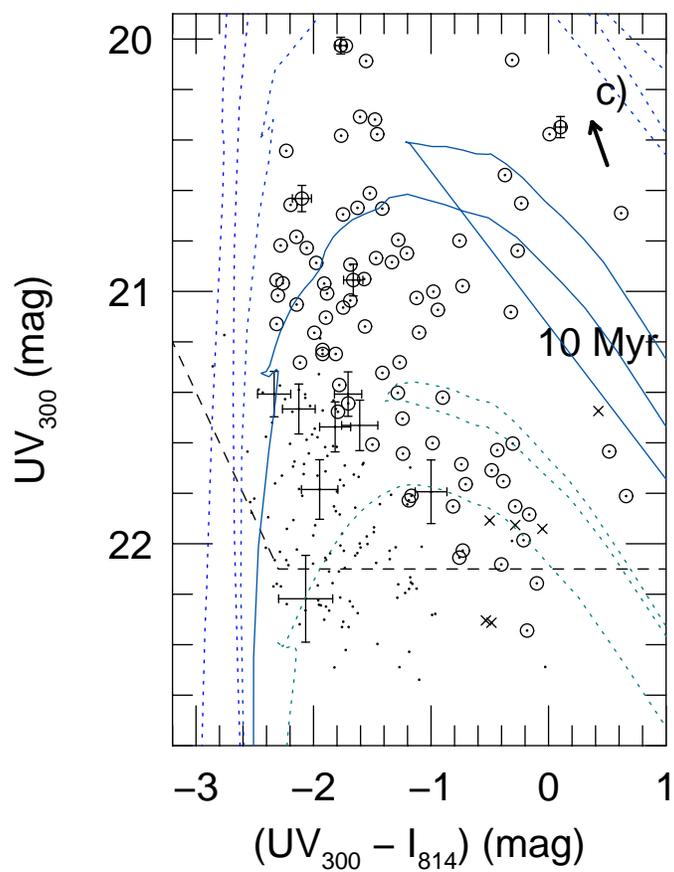
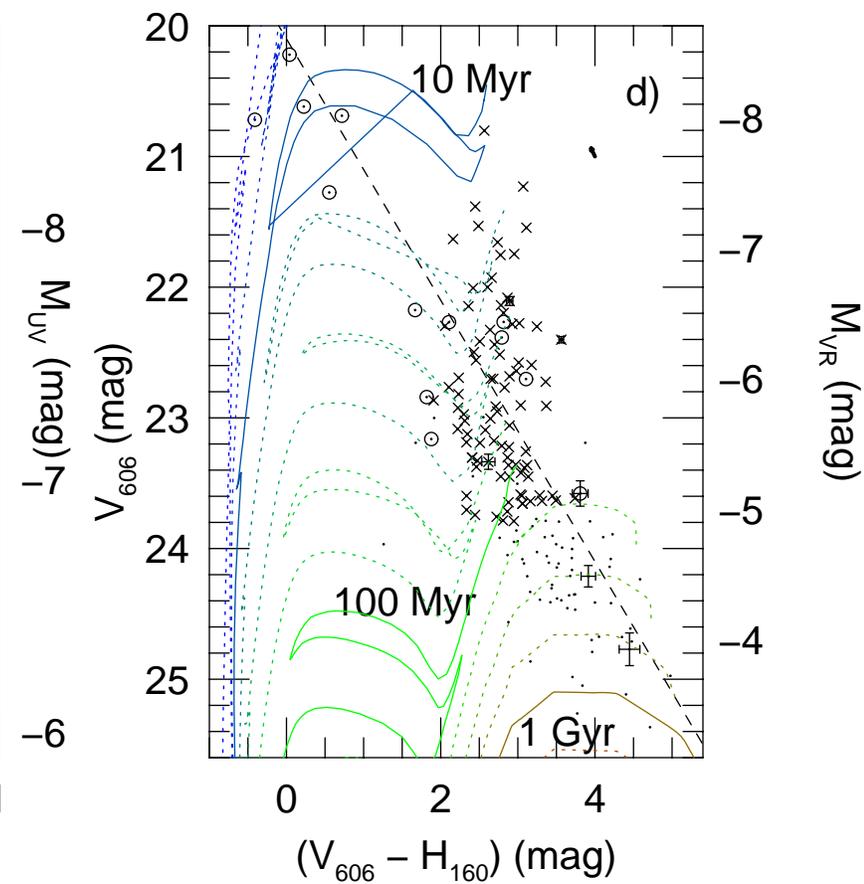
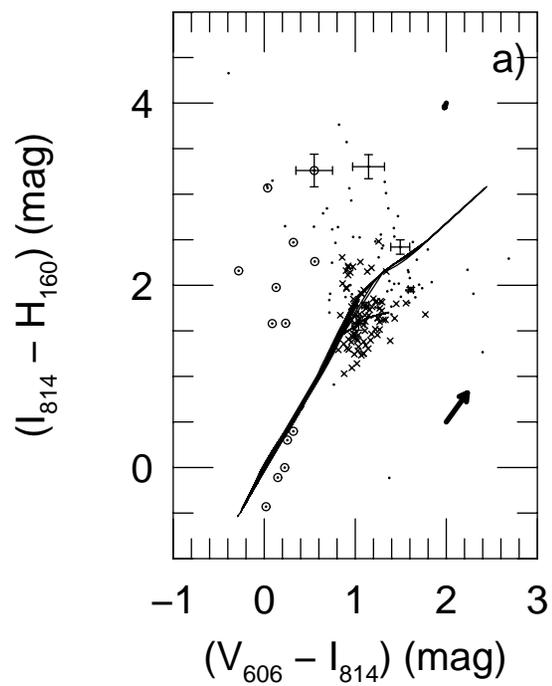
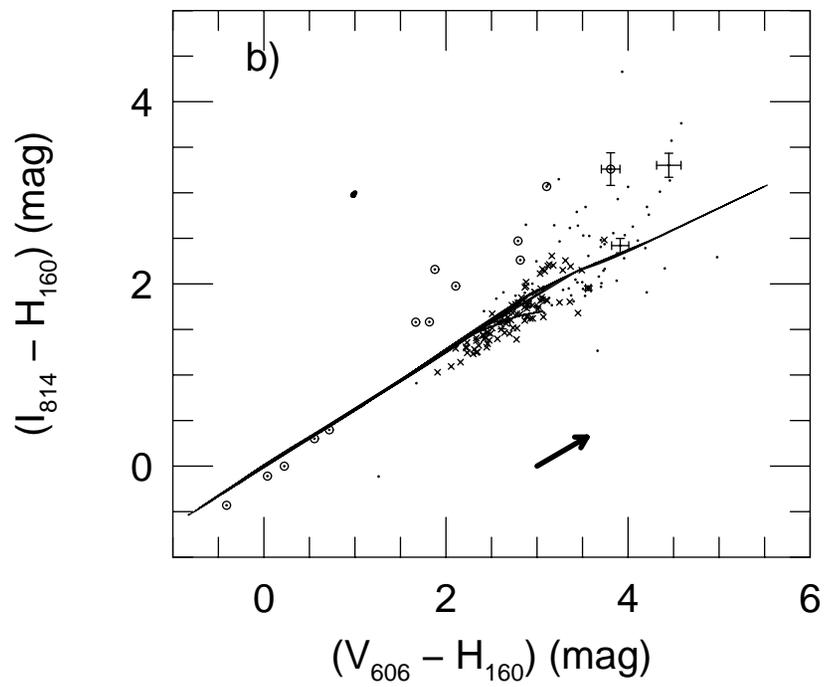
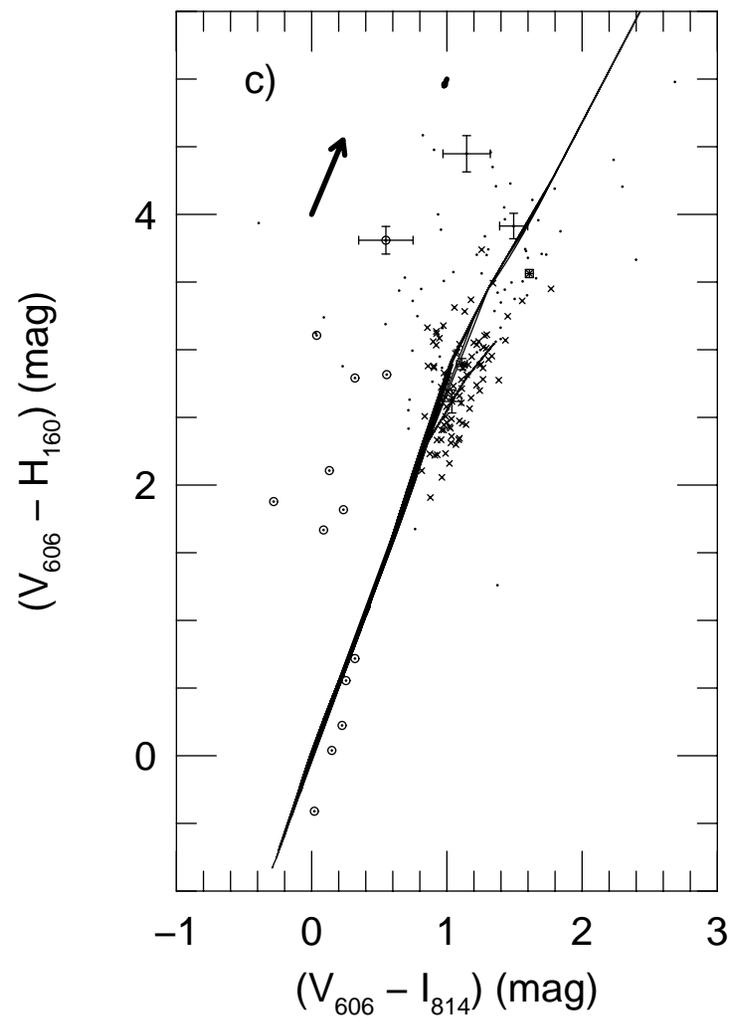
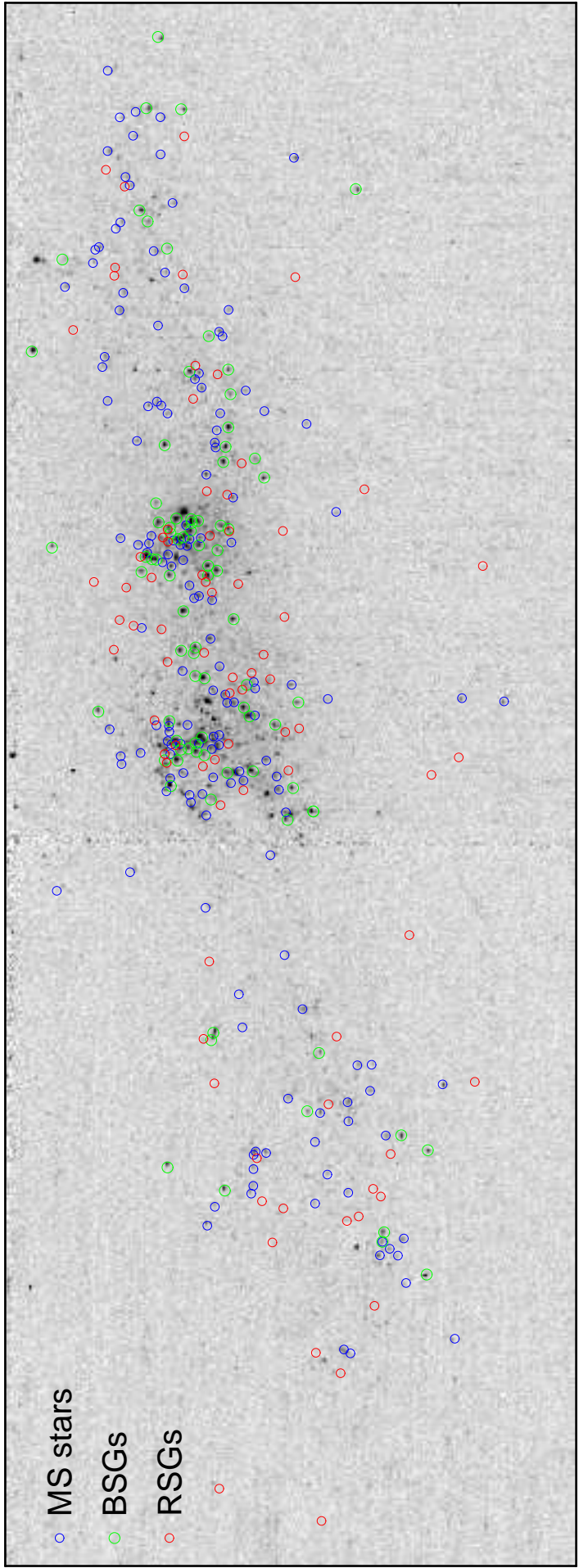
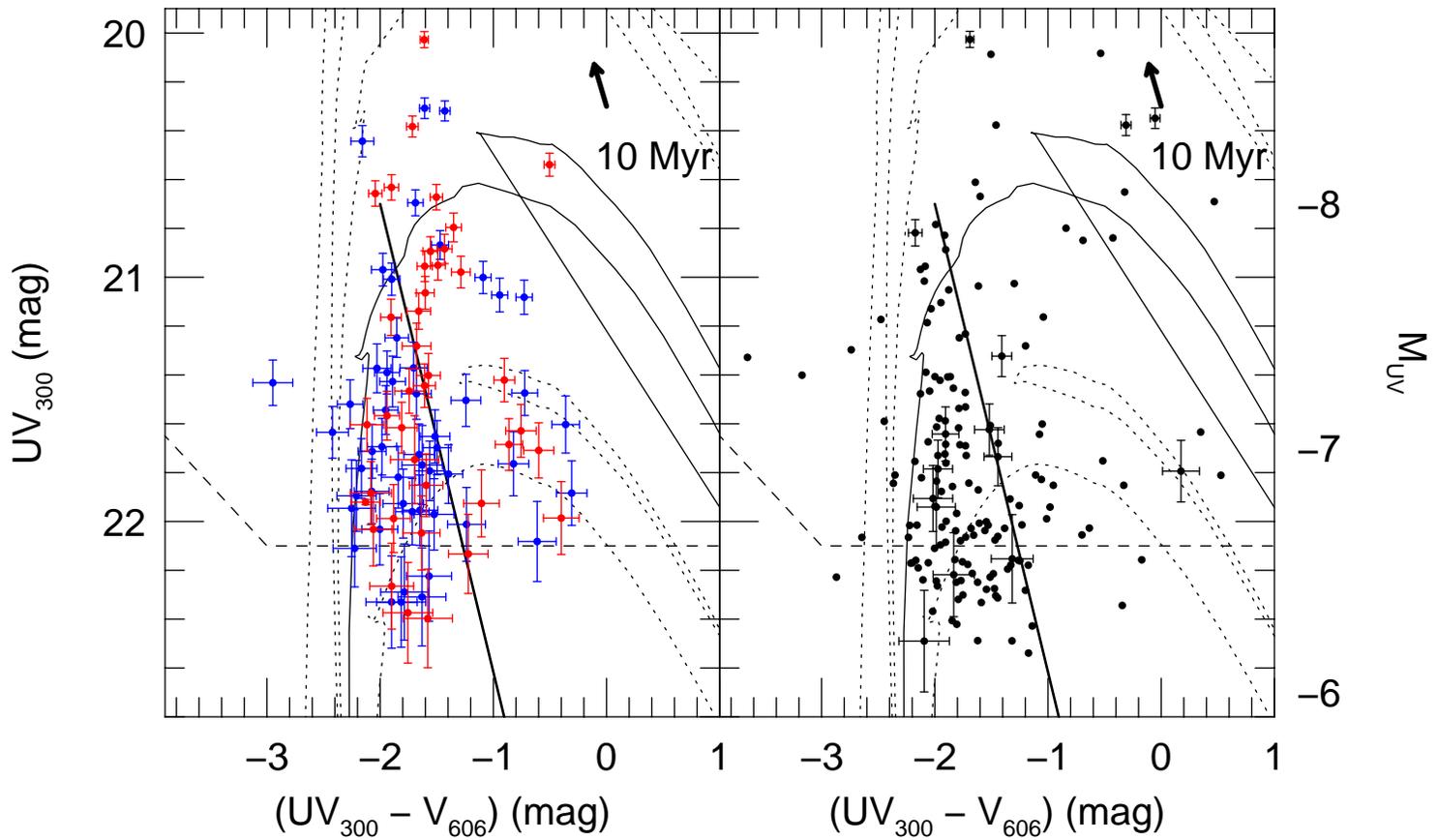
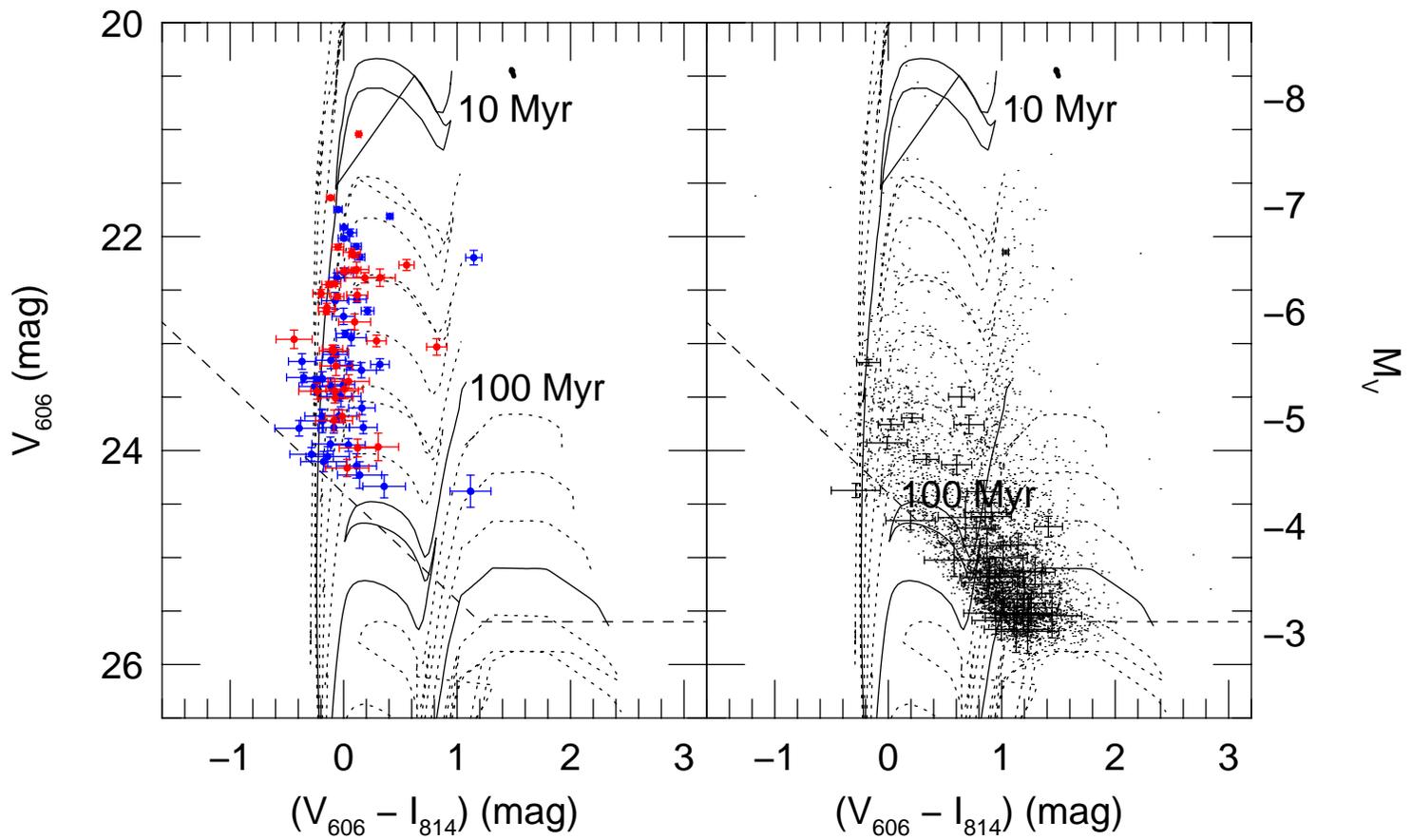
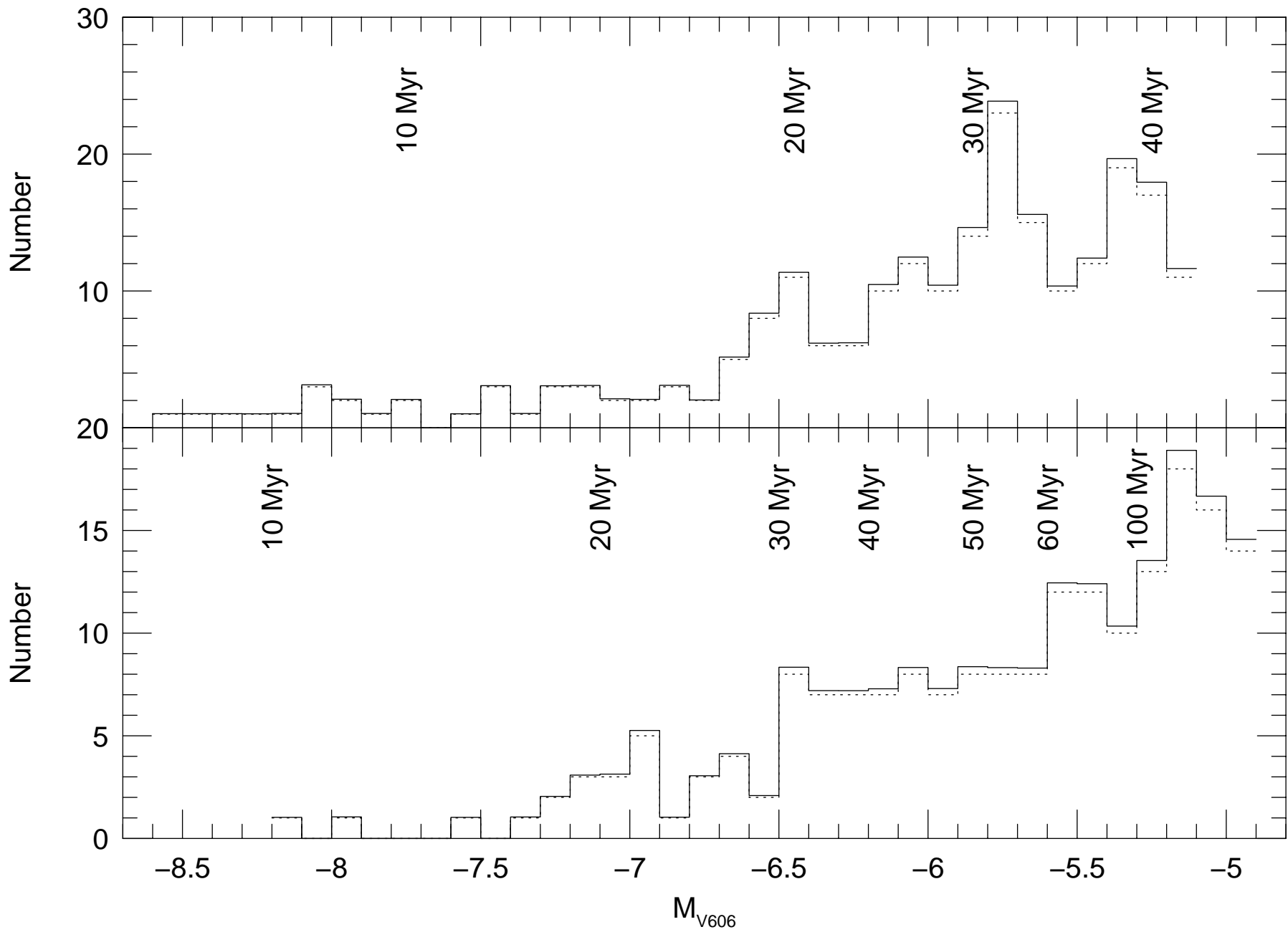
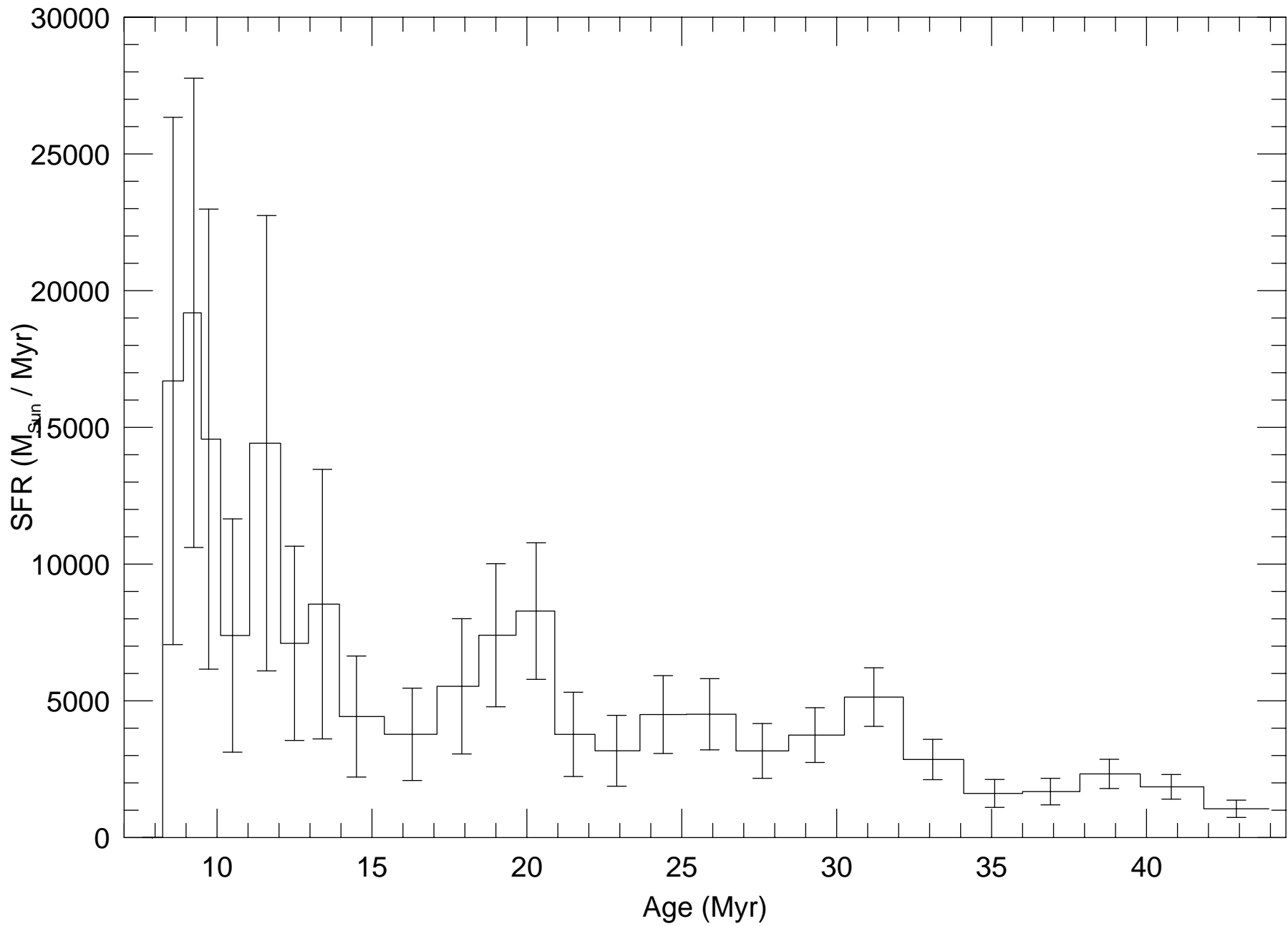
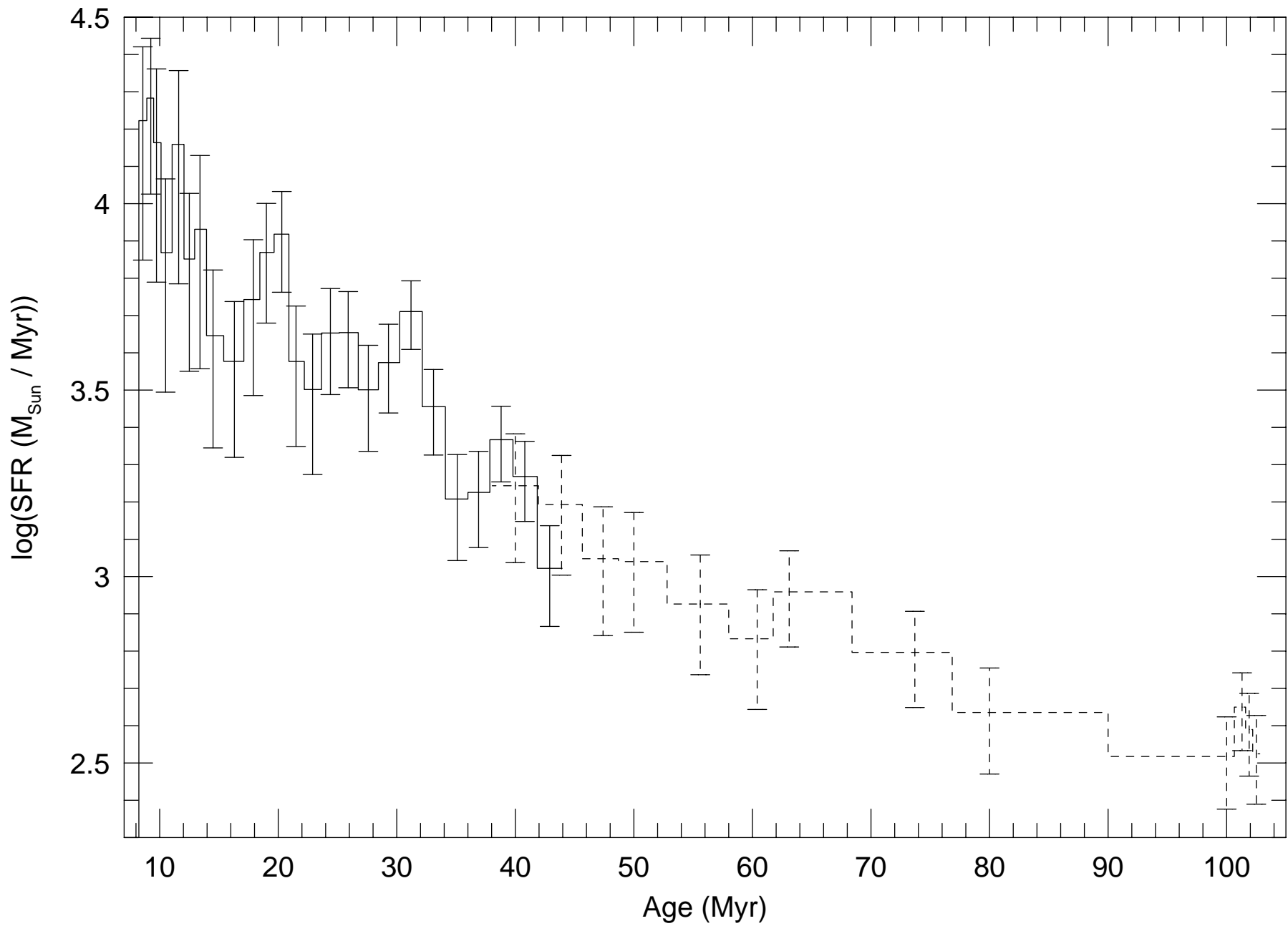
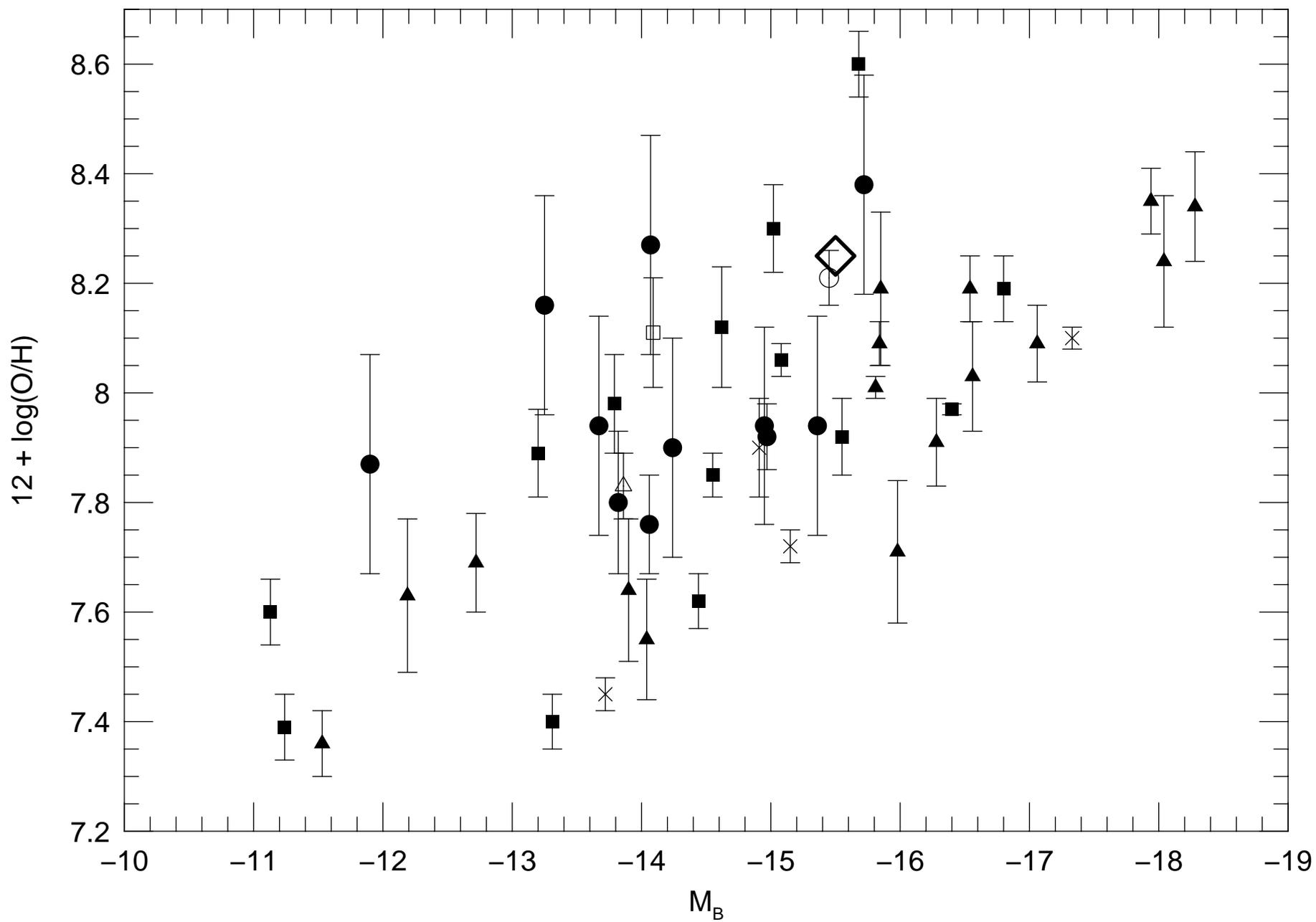
\end{document}